\documentclass[11pt,a4paper]{article}
\usepackage{mystyle}
\usepackage{soul}
\usepackage{graphicx}
\usepackage{dcolumn}
\usepackage{bm}

\usepackage{amsmath,amssymb,dsfont}
\usepackage{hyperref}
\usepackage{graphicx}
\usepackage{mathrsfs}
\usepackage{float}
\usepackage[normalem]{ulem}
\usepackage{braket}
\usepackage{color}
\usepackage{tikz,pgfplots}
\usepackage[final, color]{showkeys}
\definecolor{refkey}{gray}{0.75}
\definecolor{labelkey}{RGB}{155,48,48}
\renewcommand*\showkeyslabelformat[1]{%
  \fbox{\parbox[t]{0.8\marginparwidth}{\raggedright\normalfont\scriptsize\url{#1}}}}

\newcommand{\bea}{\begin{eqnarray}}
\newcommand{\eea}{\end{eqnarray}}
\newcommand{\be}{\begin{equation}}
\newcommand{\ee}{\end{equation}}
\newcommand{\ba}{\begin{align}}
\newcommand{\ea}{\end{align}}

\newcommand{\Tr}{{\rm {Tr}}}

\newcommand{\tr}{ {\rm Tr}}
\newcommand\rref[1]{(\ref{#1})}

\def\trt{\mathbb T{\rm r}}
\def\strt{\mathbb S\mathbb T{\rm r}}

\newcounter{source}
\newcommand\source[1]{%
    \tikz[remember picture,baseline,inner sep=0pt] {%
        \node [name=source-\thesource,anchor=base]{$#1$};
    }%
    \setcounter{target}{0}
    \stepcounter{source}
}

\newcounter{target}
\newcommand\target[1]{%
    \tikz[remember picture,baseline,inner sep=0pt] {%
        \node [name=target-\thetarget,anchor=base]{$#1$};
    }%
    \setcounter{source}{0}
    \stepcounter{target}%
}

\newcommand\doubleplus{{+\kern-1.0ex+\kern0.8ex}}
\newcommand*{\doubleplusB}{\mathrel{\vcenter{\baselineskip1.0ex \lineskiplimit0pt
                     \hbox{\tiny+}\hbox{\tiny+}}}%
}
\renewcommand{\ddag}{\doubleplusB}

\preprint{{\tt CERN-TH-2021-187}}
\title{Generalized Spectral Form Factors and \\  the Statistics of Heavy Operators}

 \author[a]{Alexandre Belin,}
 \author[b]{Jan de Boer,}
  \author[c]{Pranjal Nayak,}
 \author[c]{Julian Sonner}
 \affiliation[a]{\it CERN, Theory Division,\\1 Esplanade des Particules, Gen\`{e}ve 23, CH-1211, Suisse\\}
  \affiliation[a]{\it Institute for Theoretical Physics,\\University of Amsterdam, PO Box 94485, 1090 GL Amsterdam, The Netherlands\\}
  \affiliation[c]{\it D\'{e}partement de Physique Th\'{e}orique, Universit\'{e} de Gen\`{e}ve,\\24 quai Ernest-Ansermet, 1211 Gen\`{e}ve 4, Suisse}

\emailAdd{ a.belin [at] cern.ch}
\emailAdd{ j.deboer [at] uva.nl}
\emailAdd{pranjal.nayak [at] unige.ch}
\emailAdd{julian.sonner [at] unige.ch}

\abstract{
The spectral form factor is a powerful probe of quantum chaos that diagnoses the statistics of energy levels, but is blind to other features of a theory such as matrix elements of operators or OPE coefficients in conformal field theories. In this paper, we introduce generalized spectral form factors: new probes of quantum chaos sensitive to the dynamical data of a theory. These quantities can be studied using an effective theory of quantum chaos. We focus our attention on a particular combination of heavy-heavy-heavy OPE coefficients that generalizes the genus-2 partition function of two-dimensional CFTs, for which we define a spectral form factor. We probe heavy-heavy-heavy OPE coefficients and find statistical correlations that agree with the OPE Randomness Hypothesis: these coefficients have a random matrix component in the ergodic regime. The EFT of quantum chaos predicts that the genus-2 spectral form factor displays a ramp and a plateau. Our results suggest that this is a common property of generalized spectral form factors.

}

\begin{document}

\maketitle


\section{Introduction}

The chaotic nature of generic quantum many-body systems has triggered interest across many fields of physics, ranging from nuclear physics \cite{wigner1993characteristic} and condensed matter theory (see e.g. \cite{beenakker1997random}) to quantum gravity \cite{Kazakov:1985ea}. This converging interest can be explained by one of the most fascinating concepts in physics: universality. It is perhaps in quantum chaos that lies the strongest form of universality: the hamiltonian of any generic (i.e. non-integrable) quantum system behaves approximately like a random matrix \cite{bohigas1984characterization}. To be precise, it is the statistics of nearby energy levels that obey universal distributions. The most common probe of energy level statistics is the spectral form factor
\be
Z(\beta+i t) Z(\beta -it) = \sum_{n,m} e^{-\beta (E_n+E_m)} e^{it(E_n-E_m)} \,.
\ee
The behaviour of the spectral form factor is universal in chaotic systems (see for example \cite{Cotler:2016fpe,Altland:2020ccq} for a recent discussion). It decreases at early times due to the destructive nature of the interference caused by the phases. At sufficiently late times, which depends on the system under investigation, this decay gives way to erratic oscillations. However, the mean signal of this erratic curve still follows a universal pattern: it rises linearly (this is known as the \textit{ramp}) before saturating on a \textit{plateau}. This behaviour has a physical meaning: it is related to the repulsion of nearby eigenvalues, and ultimately comes from the fact that the spectrum of the Hamiltonian is discrete \cite{Altland:2017eao, Cotler:2016fpe}.

Another useful probe of chaos is the auto-correlation function
\be \label{autocor}
\braket{O(t) O(0)}_{\beta} =  \sum_{n,m} |O_{nm}|^2 e^{-\beta E_n} e^{it(E_n-E_m)} \,,
\ee 
where $O_{mn}$ are the matrix elements of the operator $O$ in energy eigenstates. The behaviour of this quantity is also universal, at least for an appropriate choice of operator $O$ \cite{Altland:2021rqn}. The time-scales can, however, be different than for the spectral form factor. Both the spectral form factor and the auto-correlation function are universal probes of chaos that can be defined in arbitrary quantum systems. There may also exist other probes of chaos that are special to certain classes of quantum systems. The goal of this paper is to explore such probes in the context of conformal fields theories (CFTs).

Conformal symmetry organizes the dynamics of the theory in a convenient manner. The spectrum of the Hamiltonian is given by the spectrum of local operators $\Delta_i$, and the rest of the dynamical data is encoded in the OPE coefficients $c_{ijk}$. At the local level, a CFT is fully specified by this data, and quantum chaos must therefore be reformulated in terms of the data $\{\Delta_i, c_{ijk}\}$. There are various quantities that can be built from this dynamical data and that would encode properties of quantum chaos. Many quantities of this type have not been studied from the point of view of RMT universality and in this work, we initiate a program to do so. Consider the quantity\footnote{In $d=2$, quantities like this can have a nice path integral interpretation on a higher genus surface. In $d>2$, this may no longer be the case, but they can be defined nonetheless.}
\be
\mathcal{Z}(\beta_1, \cdots,\beta_n) \equiv \sum_{{\Delta_i}} f\left[c \right] e^{-\sum_{i}\beta_i \Delta_i} \,,
\ee
where $f\left[c\right]$ is a function that depends on at least $2n/3$ OPE coefficients. An example with $n=6$ would be $c_{ijk}c_{ijl}^*c_{mnl}c_{mnk}^*$. We can now build a \textit{generalized spectral form factor}
\be
F(t_1, \cdots, t_2) \equiv |\mathcal{Z}(\beta_1+it_1, \cdots,\beta_n+it_n)|^2 \,.
\ee
In this paper, we present a general framework to embed the dynamical data of the CFT in a quantum mechanical setting, where we can apply random matrix theory.

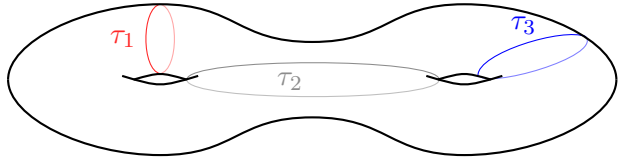
\begin{figure}[t]
\centering
\begin{tikzpicture}
	\draw[thin, blue] (7.6,0.57) .. controls (7.3,0.7) and (6.18,0.35) .. node (b) [above] {$\tau_3$}  (6.18,0.06) ;
	\draw[thin, blue!50] (7.6,0.57) .. controls (7.8,0.3) and (6.3,-0.13) .. (6.18,0.06) ;
	\draw[thin, gray!50] (5.67,0) .. controls (5.67,-0.3) and (2.35,-0.3) .. (2.35,0) ;
	\draw[thin, gray!90] (5.67,0) .. controls (5.67,0.3) and (2.35,0.3) .. node (g) [below left] {$\tau_2$} (2.35,0) ;
	\draw[thin, red!85] (2,0.08) .. controls (1.75,0.13) and (1.75,0.95) .. node (r) [left] {$\tau_1$} (2,1) ;
	\draw[thin, red!35] (2,0.08) .. controls (2.25,0.13) and (2.25,0.95) .. (2,1) ;
	\draw[thick] (0,0) .. controls (0,0.5) and (1,1) .. (2,1) .. controls (3,1) and (3,0.5) .. (4,0.5) .. controls (5,0.5) and (5,1) .. (6,1) .. controls (7,1) and (8,0.5) .. (8,0);
	\draw[thick] (0,0) .. controls (0,-0.5) and (1,-1) .. (2,-1) .. controls (3,-1) and (3,-0.5) .. (4,-0.5) .. controls (5,-0.5) and (5,-1) .. (6,-1) .. controls (7,-1) and (8,-0.5) .. (8,0);
	\draw[thick] (1.5,0.05) .. controls (2,-0.1) and (2,-0.1) .. (2.5,0.05) ;
	\draw[thick] (1.65,0.001) .. controls (2,0.1) and (2,0.1) .. (2.35,0.001) ;
	\draw[thick] (5.5,0.05) .. controls (6,-0.1) and (6,-0.1) .. (6.5,0.05) ;
	\draw[thick] (5.65,0.002) .. controls (6,0.1) and (6,0.1) .. (6.35,0.002) ;
\end{tikzpicture}
\caption{Genus-2 surface with three different cycles related to the three moduli $\tau_i$.}
\label{fig.g2s}
\end{figure}
While we expect that this framework will work in full generality, on practical grounds, we focus on a simple probe of heavy OPE coefficients that is inspired by the genus-2 partition function in two-dimensional CFTs. This is defined as
\be \label{genus2pf}
Z(\tau_1,\tau_2,\tau_3) = \sum_{O_1,O_2,O_3} |c_{123}|^2 q_1^{\Delta_1} q_2^{\Delta_1} q_3^{\Delta_3} \,, \qquad q_i=e^{2\pi i \tau_i} \,.
\ee
The sum over triplets of operators is a sum over \textit{all} operators of the CFT. One could naturally reduce the sums to be only over primaries by defining appropriate conformal blocks. For simplicity, we will not do so in this paper as we work in a toy-model which neglects the contribution of descendants. We comment on this further below.

For concreteness, it is useful to view \rref{genus2pf} in the context of two-dimensional CFTs, but the calculations we perform are more general. For $d=2$, the $\tau_i$ are moduli of the genus-2 surface and are related to its period matrix, see Fig. \ref{fig.g2s}. This quantity is then given by a tripled sum over energy eigenstates, weighted by the OPE coefficients squared (see  \cite{Belin:2017nze,Cardy:2017qhl,Cho:2017fzo,Collier:2019weq} for various studies of the genus-2 partition function). In CFTs, $O_{mn}=c_{Onm}$, such that one might think that this quantity is similar to the auto-correlation function: a Boltzmann sum weighted by OPE coefficients. However, there is a crucial difference. The expression \rref{genus2pf} does not correspond to the expectation value of a fixed operator; instead, it is like a correlation function where the probe operators are summed over. Moreover, in the small $\tau_i$ limit, the sum is dominated by OPE coefficients with three heavy operators, such that it probes different physics than \rref{autocor}.

We now introduce the associated \textit{genus-2 spectral form factor} as a new probe of quantum chaos. It is defined as
\be \label{genus2sffintro}
F(t_1,t_2,t_3)= Z(\tau_1+it_1,\tau_2+it_2,\tau_3+it_3) Z(\tau_1+it_1,\tau_2+it_2,\tau_3+it_3)^* \,.
\ee
This quantity is sensitive to both the spectral statistics (it involves six densities of states) and the statistics of OPE coefficients. One would like to study the behaviour of this quantity in chaotic (i.e. non-integrable) CFTs. 
Unfortunately, an explicit computation of this quantity appears to be difficult. It is not known how to solve non-integrable CFTs analytically and the conformal bootstrap is not presently powerful enough to efficiently probe a quantity like \rref{genus2sffintro} numerically. Instead, we investigate the implications of random matrix universality on the genus-2 spectral form factor.  We propose an alternative description of the genus-2 spectral form factor that we expect to be valid at sufficiently late times and proceed to study it using insights from random matrix theory. Therefore, we can regard our random-matrix analysis as a toy-model that closely captures the relevant late time physics.
\begin{figure}[th]
\centering
\begin{tikzpicture}
	\draw[thick,->] (0,0.5) node(orgn){} -- (0,5) node[anchor=south east] {$\ln|Z|^2$};
	\draw[thick,->] (0,0.5) -- (6.0,0.5) node[anchor=north west] {$\ln t$};
	\fill[fill=blue!20] (1,1)--(2.5,2.6)--(2.5,2.3);
	\fill[fill=blue!20] (2.5,2.3) rectangle (5.0,2.6);
	\draw [blue!25,
	declare function={
	fn1(\t,\b)= 1 + ln(1+ \t + 0.3*\t*(sin(\b*\t)) + 0.3*\t*(sin(\b*random(0.5,2)*\t)));
	fnx(\x) = 1+ln(1+\x);
	}]
	plot [domain=0:1.5, samples=144, smooth] ({fnx(\x)},{fn1(\x,2500)});
	\draw [blue!25,
	declare function={
	fn1(\t,\b)= 1 + ln(1+ \t + 0.3*\t*(sin(\b*\t)) + 0.3*\t*(sin(\b*random(0.5,2)*\t)));
	fnx(\x) = 1+ln(1+\x);
	}]
	plot [domain=1.5:3.3, samples=144, smooth] ({fnx(\x)},{fn1(\x,2000)});
	\draw [blue!25,
	declare function={
	fn2(\t,\b)= 2.5 + ln(1 + 0.3*(sin(\b*\t)) + 0.3*(sin(\b*random(0.5,2)*\t)));
	fnx2(\x) = 2.5+ln(1+\x);
	}]
	plot [domain=0:1, samples=74, smooth] ({fnx2(\x)},{fn2(\x,5000)});
	\draw [blue!25,
	declare function={
	fn2(\t,\b)= 2.5 + ln(1 + 0.3*(sin(\b*\t)) + 0.3*(sin(\b*random(0.5,2)*\t)));
	fnx2(\x) = 2.5+ln(1+\x);
	}]
	plot [domain=1:1.6, samples=74, smooth] ({fnx2(\x)},{fn2(\x,5000)});
	\draw [blue!25,
	declare function={
	fn2(\t,\b)= 2.5 + ln(1 + 0.3*(sin(\b*\t)) + 0.3*(sin(\b*random(0.5,2)*\t)));
	fnx2(\x) = 3.3+ln(1+\x);
	}]
	plot [domain=0:1, samples=74, smooth] ({fnx2(\x)},{fn2(\x,5000)});
	\draw [blue!25,
	declare function={
	fn2(\t,\b)= 2.5 + ln(1 + 0.3*(sin(\b*\t)) + 0.3*(sin(\b*random(0.5,2)*\t)));
	fnx2(\x) = 3.7+ln(1+\x);
	}]
	plot [domain=0:1, samples=74, smooth] ({fnx2(\x)},{fn2(\x,5000)});
	\draw [blue!25,
	declare function={
	fn2(\t,\b)= 2.5 + ln(1 + 0.3*(sin(\b*\t)) + 0.3*(sin(\b*random(0.5,2)*\t)));
	fnx2(\x) = 4.1+ln(1+\x);
	}]
	plot [domain=0:1, samples=74, smooth] ({fnx2(\x)},{fn2(\x,5000)});
	\draw [blue!25,
	declare function={
	fn2(\t,\b)= 2.5 + ln(1 + 0.3*(sin(\b*\t)) + 0.3*(sin(\b*random(0.5,2)*\t)));
	fnx2(\x) = 4.5+ln(1+\x);
	}]
	plot [domain=0:1, samples=74, smooth] ({fnx2(\x)},{fn2(\x,5000)});
		\draw [blue!25,
	declare function={
	fn1(\t,\b)= 1 + ln(1+ \t + 0.3*\t*(sin(\b*\t)) + 0.3*\t*(sin(\b*random(0.5,2)*\t)));
	fnx(\x) = 1+ln(1+\x);
	}]
	plot [domain=0:1.5, samples=144, smooth] ({fnx(\x)},{fn1(\x,1863)});
	\draw [blue!25,
	declare function={
	fn1(\t,\b)= 1 + ln(1+ \t + 0.3*\t*(sin(\b*\t)) + 0.3*\t*(sin(\b*random(0.5,2)*\t)));
	fnx(\x) = 1+ln(1+\x);
	}]
	plot [domain=1.5:3.3, samples=144, smooth] ({fnx(\x)},{fn1(\x,1732)});
	\draw [blue!25,
	declare function={
	fn2(\t,\b)= 2.5 + ln(1 + 0.3*(sin(\b*\t)) + 0.3*(sin(\b*random(0.5,2)*\t)));
	fnx2(\x) = 2.5+ln(1+\x);
	}]
	plot [domain=0:1, samples=74, smooth] ({fnx2(\x)},{fn2(\x,3282)});
	\draw [blue!25,
	declare function={
	fn2(\t,\b)= 2.5 + ln(1 + 0.3*(sin(\b*\t)) + 0.3*(sin(\b*random(0.5,2)*\t)));
	fnx2(\x) = 2.5+ln(1+\x);
	}]
	plot [domain=1:1.6, samples=74, smooth] ({fnx2(\x)},{fn2(\x,2957)});
	\draw [blue!25,
	declare function={
	fn2(\t,\b)= 2.5 + ln(1 + 0.3*(sin(\b*\t)) + 0.3*(sin(\b*random(0.5,2)*\t)));
	fnx2(\x) = 3.3+ln(1+\x);
	}]
	plot [domain=0:1, samples=74, smooth] ({fnx2(\x)},{fn2(\x,4829)});
	\draw [blue!25,
	declare function={
	fn2(\t,\b)= 2.5 + ln(1 + 0.3*(sin(\b*\t)) + 0.3*(sin(\b*random(0.5,2)*\t)));
	fnx2(\x) = 3.7+ln(1+\x);
	}]
	plot [domain=0:1, samples=74, smooth] ({fnx2(\x)},{fn2(\x,2839)});
	\draw [blue!25,
	declare function={
	fn2(\t,\b)= 2.5 + ln(1 + 0.3*(sin(\b*\t)) + 0.3*(sin(\b*random(0.5,2)*\t)));
	fnx2(\x) = 4.1+ln(1+\x);
	}]
	plot [domain=0:1, samples=74, smooth] ({fnx2(\x)},{fn2(\x,2930)});
	\draw [blue!25,
	declare function={
	fn2(\t,\b)= 2.5 + ln(1 + 0.3*(sin(\b*\t)) + 0.3*(sin(\b*random(0.5,2)*\t)));
	fnx2(\x) = 4.5+ln(1+\x);
	}]
	plot [domain=0:1, samples=74, smooth] ({fnx2(\x)},{fn2(\x,4837)});
	\draw[line width = 1pt, red] (0,4.5) node(strt){} ..controls (0.5,4.5) and (0,1) .. (1,1);
	\draw[line width = 1pt, blue] (1,1).. controls (1.2,1) and (1.5,1.5) .. (2.5,2.5) -- (5.2,2.5);
\end{tikzpicture}
\caption{The genus-2 spectral form factor as a function of time. After an initial decay, the signal displays a ramp and plateau at sufficiently late times. The ramp and the plateau are the average of the noisy signal. The $\sigma$-model captures the moments of this noisy signal.}
\label{fig.g2t}
\end{figure}
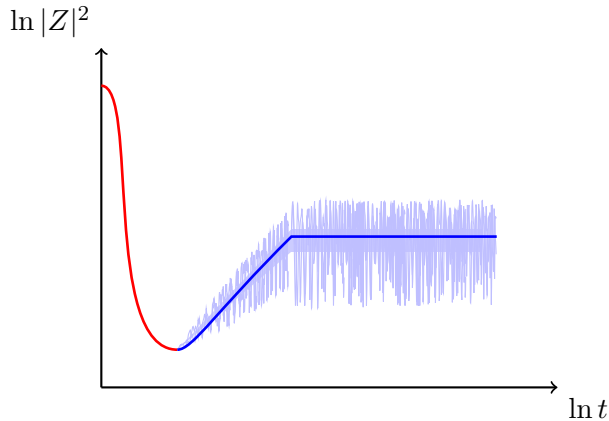

\subsection{A probe for the genus-2 partition function}
In this paper, we define an operator,
\begin{align} \label{Odefintro}
	\langle i \, j \, k |  \mathbb O  | i' \, j' \, k'\rangle = c^*_{ijk} c_{i'j'k'} \,,
\end{align}
whose matrix elements encapsulate the OPE coefficients of the theory under investigation. $\ket{i},\ket{j},\ket{k}$ are energy eigenstates such that this operator can be thought of as a linear operator acting on a tripled Hilbert space. The genus-two partition function is then defined as\footnote{Here, we take a particular slice of the genus-2 moduli space where the three moduli are the same.} 
\be\label{eq.Genus-2.Partition.Function}
Z(\tau_1=\tau_2=\tau_3=i \beta) = \tr \left[ \mathbb O e^{-\beta(H_1+H_2+H_3)} \right]\,,
\ee
which is a ``thermal" one-point function on this tripled Hilbert space. Having turned the problem into an effective quantum-mechanical one, we use insights from operator statistics and random matrix universality to probe the genus-2 spectral form factor.

In chaotic quantum systems, aspects of operator statistics are captured by the Eigenstate Thermalization Hypothesis (ETH) \cite{deutsch1991quantum,srednicki1994chaos}. For simple operators $O$, the matrix elements have the general structure
\be
\braket{n|O|m}= \delta_{mn} f(\bar{E}) + g(\bar{E},\delta E) e^{-S(\bar{E})/2} R_{mn} \,,
\ee
where $\bar{E}$ and $\delta E$ are the mean energy and energy difference, and $f$ and $g$ are two smooth functions related to the microcanonical one- and two-point functions. The coefficients $R_{mn}$ are fixed in any given theory, but their statistical distribution is approximately gaussian random. For this reason, we call them pseudo-random variables. The ETH ansatz has not been proven, even though it can be verified numerically, for example in the holographically relevant case of the SYK model \cite{Sonner:2017hxc}. Analytical results pointing to the validity of ETH in this model have furthermore been established in \cite{Nayak:2019khe, Nayak:2019evx}. ETH has also been studied in higher dimensions, \cite{Lashkari:2016vgj,  Dymarsky:2018ccu,Delacretaz:2020nit}, in particular in 2-dimensional CFTs where Virasoro symmetry introduces additional subtleties \cite{Fitzpatrick:2015zha, Basu:2017kzo, Romero-Bermudez:2018dim, Brehm:2018ipf, Hikida:2018khg, Guo:2018pvi, Maloney:2018yrz, Dymarsky:2018lhf, Dymarsky:2019etq, Datta:2019jeo, Besken:2019bsu}.

More generally, there is a symmetry-breaking argument (causal symmetry breaking) that describes the fine-grained structure of the energy spectrum as well as the associated eigenstates of a chaotic many-body system \cite{Altland:2020ccq}. The resulting effective field theory dictates both spectral and operator correlations (see Section \ref{sec.basics}). For the statistics of operator correlations in chaotic systems, in addition to random-matrix like correlations of energy eigenvalues, Haar unitary averages over the eigenstates contribute equally importantly \cite{Altland:2021rqn}. This lends credibility to a typicality argument in favor of the ETH \cite{deutsch1991quantum,srednicki1994chaos}: if we replace an energy eigenstate with a Haar-random state in a microcanonical window, a version of ETH follows. The jump to ETH is to assume that the energy eigenstates are as good as Haar-random states in a chaotic system since the Hamiltonian is close to a random matrix.\footnote{It is worthwhile to note that ETH also describes the physics of thermalization, which occurs at much earlier time scales than the random matrix phase. }

For chaotic CFTs, a generalization of ETH called the OPE Randomness Hypothesis (ORH) was proposed in \cite{Belin:2020hea}. It proposes to treat the OPE coefficients involving heavy (i.e. $\Delta\to\infty$) operators as pseudo-random variables. This includes the ETH type OPE coefficients $O_{mn}=c_{Omn}$ for some light operator $O$, but also makes a prediction for the structure of other types of OPE coefficients, in particular the OPE coefficients of three heavy operators. One of the aims of this paper is to provide evidence for the ORH using Haar-typicality and  random matrix universality.

An important motivation behind the present work is also to illuminate our understanding of the AdS/CFT correspondence. In the dual gravitational theory, the semi-classical limit was long thought to only capture the early time decaying behaviour of physical observables like the auto-correlation, which ultimately leads to information loss \cite{Maldacena:2001kr}. Recent work for the nAdS${}_2$/nCFT${}_1$ correspondence between JT gravity and its matrix model dual demonstrated that the contribution of certain gravitational configurations known as Euclidean wormholes are capable of reproducing the linearly growing ramp-like behaviour \cite{Saad:2018bqo, Saad:2019lba}. A universal description in \cite{Altland:2020ccq} shows that the Euclidean wormholes capture the moments of the signals at sufficiently late times and hence reproduce the ramp-plateau behaviour. In higher dimensions, a detailed understanding of this picture is still being developed (see \cite{Cotler:2020ugk,Belin:2020hea} for gravity in AdS$_3$). In particular, it is not known what gravitational configurations reproduce the plateau behaviour. From \cite{Altland:2021rqn}, we have strong indications that the spacetime topology of these configurations should be the same as that of the wormhole solutions that explain the ramp.

From a wormhole perspective, the 2d genus-2 spectral form factor is an ideal observable to sharpen our understanding of holography. On the gravitational side, the relevant wormhole is called the genus-2 wormhole, stretching between two asymptotic genus-2 surfaces. Unlike the geometries relevant to compute the standard spectral form factor, the genus-2 wormhole is a true saddle of the gravitational equations of motion, and it can be perturbatively stable when embedded in certain top-down theories \cite{Maldacena:2004rf}. Precisely matching the expressions that we derive in this paper with the gravitational contribution of the genus-2 wormhole requires explicit knowledge of the statistical distribution of heavy-heavy-heavy OPE coefficients. Asymptotic formulas for these OPE coefficients are known, either for the Gaussian part \cite{Cardy:2017qhl,Collier:2019weq} or for the non-gaussian corrections \cite{Belin:2021ryy}. However, one can only trust that these asymptotic formulas correctly encode the statistical distribution of the OPE coefficients if one can argue that they are pseudo-random variables. The asymptotic formulas are blind to his fact, and one of the main goals of this paper is to explain to what extent the OPE coefficients have a random nature to them.


\subsection{Summary of Results}

The two main results we present in this paper are as follows. First, we provide evidence for the ORH based on random matrix theory. We show that the operator \rref{Odefintro} satisfies the following statistics. Its mean is given by\footnote{The overline notation $\overline{\phantom{a}\cdot\phantom{a}}$ means that we have effectively evaluated the matrix elements in random matrix theory. Thus, the mean and average we obtain are valid in the ergodic limit, so these results should hold for nearby matrix elements (defined by the Thouless time). Whether or not this extends to the full microcanonical window is a more difficult question that cannot be probed from ergodicity. Similar statements apply to the ETH (see for example \cite{Dymarsky:2018ccu,Richter:2020bkf}).}
\begin{align}
\label{eq.signal-same.intro}
	\overline{\langle i_i i_2 i_3|\mathbb O |i'_1 i'_2 i'_3\rangle} &= \trt \mathbb{O}  ~ \frac1{D_1D_2D_3} ~ \delta_{i_1,i_1'}\delta_{i_2,i_2'}\delta_{i_3,i_3'} ~,
\end{align}
and its variance is
\bea\label{eq.variance.intro}
	\overline{\langle i_i i_2 i_3|\mathbb O |i'_1 i'_2 i'_3\rangle^2}- \overline{\langle i_i i_2 i_3|\mathbb O |i'_1 i'_2 i'_3\rangle}^2 ~ &\approx& ~ \left(\frac{1}{D_1 D_2 D_3}\right)^2 \Big[\ ~\trt \mathbb{O} ~ \trt \mathbb{O}  \\
	&\hspace{-5.75cm}+&\hspace{-3cm} \trt_{\rm sunset} \mathbb{O}^2\Big(\delta_{i_1i'_1}\delta_{i_2i'_2} + \delta_{i_1i'_1}\delta_{i_3i'_3} + \delta_{i_3i'_3}\delta_{i_2i'_2} ~+ \delta_{i_1i'_1} + \delta_{i_2i'_2} + \delta_{i_3i'_3}\Big)\Big]~ \,. \notag
\eea
Here, $D_i$  are the dimensions of the three (microcanonical) Hilbert spaces which we have taken to be distinct for the three energy windows $i_1,i_2,i_3$,\footnote{More complicated expressions where all three windows are the same can be found in Section \ref{singlewindsec}.}  and $\trt \mathbb{O}$ and $\trt_{\rm sunset} \mathbb{O}^2$ represent microcanonical traces. The sunset trace is a particular contraction given by  $c_{ijk}c^{*}_{ijl}c_{mnl}c_{mnk}^{*}$.

The mean and variance of the matrix elements of $ \mathbb{O}$ are compatible with the OPE coefficients $c_{ijk}$ being random and approximately Gaussian random variables, as advocated by the OPE Randomness Hypothesis. For the distribution to be approximately Gaussian,  $\trt_{\rm sunset} \mathbb{O}^2$ must be exponentially suppressed compared to $(\trt \mathbb{O})^2$. Here, this suppression is not built-in and must be taken as an additional assumption.\footnote{A similar assumption enters in ETH, where the microcanonical one- and two-point functions are assumed to be of the same order.} Note however that in two-dimensional CFTs, the sunset contraction has indeed be shown to be exponentially suppressed \cite{Belin:2021ryy}.

It is tempting to assemble the mean and variance of $ \mathbb{O}$ into a statistical formula with random variables as in ETH, which we do in \rref{resultdiffwindow2}. There are however crucial points to be kept in mind. The most important caveat is that the diagonal and random components of $c_{ijk}c_{lmn}^*$ are the same size. This is in fact expected from writing down a statistical ansatz for a composite structure such as $c_{ijk}c_{lmn}^*$. The composite nature of this object gives its variance a size comparable to its mean, which is expected from cross-Wick contractions that appear in the calculation of the variance. This is very similar to correlation functions of multi-trace operators in a generalized free theory, which contain connected components even though the underlying theory is Gaussian.

Our second result is the late-time behaviour of the genus-2 spectral form factor. It is most easily written in energy space, in terms of the resolvent (see \rref{eq.new-resolvent}). We find that the genus-2 spectral form factor resolvent is given by 
\begin{equation}
R(s)=	\frac{2\pi^2\rho^2(E)}{D^6} \left(\pi \delta(s) + 1 - \frac{\sin^2(s)}{s^2}\right) \tr (\mathbb O)^2  ~,
\end{equation}
where $s$ is the energy difference in units of mean level spacing of the tripled Hilbert space. This is the genus-2 version of the standard sine-kernel. The sin kernel is well-known for the standard spectral form factor, and predicts both a ramp and a plateau. We see that the behaviour is similar for the genus-2 spectral form factor. Here, it appears that spectral correlations are the dominant effect and the operator statistics do not really play a prominent role (this is why no sunset trace appears). Note however that the Thouless time, i.e. the time at which we can trust the random matrix behaviour may depend on the observable, and thus the correlations of the OPE coefficients can be important in determining the Thouless time for the genus-2 spectral form factor. Note that in any case, the ramp time which is of the order of inverse mean level spacing, is much larger here because of the enlarged dimensionality of the tripled Hilbert space.

The main assumption that we make in this paper is that random matrix universality applies to conformal field theories, i.e. that CFTs have a finite Thouless time. While this statement is expected to hold in arbitrary quantum systems (in fact at the Thouless time, even the locality of the quantum field theory has been washed away), it has not been proven and the Thouless time needs to be calculated theory by theory. It remains one of the big open problems to understand quantum chaos in QFTs and CFTs, and goes beyond the scope of this work.

This paper is organized as follows. In section \ref{sec.basics}, we review the EFT of quantum chaos and introduce the quantum mechanical model for the OPE coefficients in terms of a linear operator on a tripled Hilbert space. In section \ref{sec.ope}, we study Haar-averages over unitaries and determine the statistics of OPE coefficients making use of the linear operator $\mathbb O$. In section \ref{sec.g2}, we use the EFT of quantum chaos to study the genus-2 spectral form factor. We end with a conclusion and a discussion of open questions in section \ref{sec.discussion}.


\section{Our Setup and the EFT of Quantum Chaos}
\label{sec.basics}
In this section, we start with a review of the effective field theory that describes the ergodic regime of chaotic quantum systems \cite{Altland:2020ccq}. The quantum ergodic regime of a physical theory is defined as the one in which the behaviour of physical observables is indistinguishable from that in a random matrix theory (RMT).\footnote{The exact nature of which RMT itself depends on the time-reversal symmetry and the fermionic symmetries of the physical theory. This is classified in a 10-fold way by Altland and Zirnbauer, \cite{PhysRevB.55.1142}, building on the pioneering work of Wigner, Dyson and Mehta \cite{wigner1993characteristic,dyson1962statistical,mehta-book}.} More accurately, one should say that the behaviour is indistinguishable from a typical representative of the random matrix ensemble. Hence, it is possible to model a physical system using a random matrix ensemble in this limit. We then introduce a new formalism that is well suited to study the behaviour of OPE coefficients and the genus-2 partition function in the quantum ergodic limit. We will describe how to translate heavy-heavy-heavy OPE coefficients in terms of a standard quantum mechanical system. In subsequent sections, we use this formalism to probe the statistical behaviour of OPE coefficients and the genus-2 spectral form factor.


\subsection{The effective field theory of quantum chaos}
\label{sec.review}
Here, we give a brief overview of our main tool for studying the spectral probes in quantum theories: a symmetry-based effective field theory of quantum chaos described in \cite{Altland:2020ccq}. We refer the reader to \cite{Altland:2021rqn} for a more detailed exposition.

As we have emphasized a few times by now, chaos manifests itself in the fine structure of energy levels, and does so in a remarkably universal fashion \cite{wigner1993characteristic, dyson1962statistical, altland1997nonstandard}. One natural way to understand this universality is by identifying a symmetry and its associated breaking pattern, which gives rise to chaotic spectral correlations. Furthermore, in order to fully resolve the fine-grained structure of the energy spectrum and associated eigenfunctions, it turns out to be essential to work in a framework that is capable of controlling both perturbative (in $e^S$) {\it and} non-perturbative (in $e^S$) contributions to our quantities of interest. It is thus fortuitous that both these aims can be achieved by considering ratios of so-called spectral determinants, the first one of interest to the present discussion being
\be\label{eq.Z4GenFunctional}
{\cal Z}^{(4)} \left( z_1\,\ldots ,z_4   \right) = \frac{{\rm Det} \left(z_1-H \right) {\rm Det} \left(z_2-H \right)}{{\rm Det} \left(z_3-H \right){\rm Det} \left(z_4-H \right)}
\ee
where $H$ is the Hamiltonian of the chaotic quantum system in question, and $z_i$ are four different energies. As we will see in more detail below, the framework we are about to lay out here allows us to evaluate the spectral determinants above in terms of an effective field theory that becomes semi-classical for large $D = e^S$. Within this EFT we can then understand non-perturbative (in $e^S$) physics via the sum over saddle points, while perturbative (in $e^S$) physics comes from the perturbation theory around each individual saddle.

The utility of quantities like ${\cal Z}^{(4)}(\hat z)$  is that they allow us to generate insertions of the spectral density $\rho(z)$, via the relation $\mp i \pi \rho(z) = {\rm Im} \tr \frac{1}{z-H}$, where
\be
\tr \frac{1}{z-H} = \frac{\partial_z {\rm Det}(z-H)}{{\rm Det}(z'-H)} \Biggr|_{z=z'}\,.
\ee
This opens the door to study chaotic correlations of the spectral density. The main idea of the EFT of quantum chaos is to develop a field theory of this ratio of spectral determinants. It may be useful to note that this is a field theory that lives on the space of Hamiltonians, and its modes parametrize how close a given Hamiltonian is to being an RMT Hamiltonian. The crucial point, however, is the realization that such a theory necessarily possesses a continuous symmetry that reflects the fact that the four copies of the system Hamiltonian that enter \rref{eq.Z4GenFunctional} are by design identical. In order to make this symmetry apparent, as is described in detail in \cite{Altland:2020ccq} it is convenient to exponentiate the determinants via Gaussian integrals over bosonic/fermionic auxiliary variables (for determinants in denominator/numerator, respectively). The exact symmetry breaking pattern differs from one Altland-Zirnbauer symmetry class to another, but in the simplest case (the unitary class), it corresponds to a breaking of 
\be\label{eq.SymmetryBreaking}
G={\rm U}(2,2 | 2) \longrightarrow K={\rm U}(1|1)\times {\rm U}(1|1)\,,
\ee
and the EFT of quantum chaos takes the form of a non-linear sigma model with target space $G/K$. The precise sense in which one should understand this symmetry breaking is via a mean-field solution of the field theory obtained by exponentiating our ratio of determinants
\begin{equation}\label{eq.DensityEFT}
 \int [dQ] e^{-S[Q]}  :=\left\langle \frac{{\rm Det} \left(z_1-H \right) {\rm Det} \left(z_2-H \right)}{{\rm Det} \left(z_3-H \right){\rm Det} \left(z_4-H \right)} \right\rangle_{\Delta E \,\, {\rm or}\,\, P(H), 
 \ldots}\,,
\end{equation}
where $Q$ is a collective degree of freedom parametrizing the Goldstone manifold \rref{eq.SymmetryBreaking} and which is defined in detail in \cite{Altland:2020ccq}. We indicated with angle brackets that the determinants are to be averaged over small (microcanonical) energy windows, or alternatively over a probability distribution of the Hamiltonian, or any other generic averaging procedures. The main point here is that the resulting physics is dictated by our symmetry breaking principle and does therefore not depend on the averaging procedure. It is then interesting to note that such a (slightly averaged) version of a spectral determinant can be defined for an \emph{individual} quantum system, for example via averaging over some small microcanonical energy window. In other words, from the EFT point of view, an individual quantum system gives rise to exactly the same IR EFT of quantum chaos as a full random-matrix ensemble. 

This is, of course, nothing but a technical way of restating random-matrix universality: {\it an individual quantum chaotic system has spectral correlations that are well described by random matrix theory.}\footnote{Some work where the emergence of the ensembles has been discussed are: \cite{Anninos:2016szt, Balasubramanian:2020lux,Verlinde:2021kgt, Verlinde:2021jwu}.} This fundamental fact leaves its imprint on a variety of physical quantities. Admittedly, the EFT has its limitations and doesn't provide a description of arbitrary fine-grained observables. For this reason, while at late times, the EFT captures the moments of the noisy signal, it remains an open question to understand how far one can push this description to capture the entire signal. Moreover, the EFT also breaks down at low temperatures when the thermal sum is dominated by the contribution of the low energy states (those near the ground state).

In this paper, we are primarily concerned with chaotic conformal field theories, and in particular, the statistical distribution of their OPE coefficients. We are thus interested in applying the EFT approach outlined above to the operator \rref{eq.OPEoperator}. The main novelty is that we find it convenient to consider not just the energy density itself, but an ``operator-weighted" energy density, that is
\begin{equation}\label{eq.OperatorWeightedDensity}
\rho_{ O} (E) := {\rm Im} \tr \left[ \frac{1}{E-H} O \right]~.
\end{equation}
It may already be obvious to the reader at this point that such a quantity gives us a way to define spectral sums weighted by operator matrix elements. In particular, with our purpose-built operator  $\mathbb O$, the spectral sums are weighted by OPE coefficients: we precisely get the kind of quantities introduced in \rref{eq.Genus-2.Partition.Function}.
 Furthermore, this is but a small extension of the ideas we just reviewed, and proceeds as follows. We start again by noting that
\begin{equation}
\Tr \left(  \frac{1}{z-H} O \right) =\frac{\partial}{\partial h}  \frac{{\rm Det}(z-H+ h\, O)}{{\rm Det}(z-H + h' O)} \Biggr|_{h=h'=0}\,.
\end{equation}
From this we can then proceed as above and define a field theory of sourced spectral determinants, starting from
\begin{equation}\label{eq.SourcedEFT}
\int [dQ] e^{-S[Q, h]} = \left\langle \frac{{\rm Det} \left(z_1-H + h_1 O \right) {\rm Det} \left(z_2-H + h_2 O \right)}{{\rm Det} \left(z_3-H +  h_3 O \right){\rm Det} \left(z_4-H+ h_4 O \right)} \right\rangle_{\Delta E\,\,\, {\rm or}\,\, P(H), }
\end{equation}
where we have denoted all four possible sources by adding the parameter $h$ to the effective action. An important difference between the effective actions \eqref{eq.DensityEFT} and \eqref{eq.SourcedEFT} is the contribution of Haar average over unitary matrices. The equation \eqref{eq.DensityEFT} is invariant under unitary transformations of the Hamiltonian, $H\to U H U^{-1}$, and therefore only depends on the eigenvalues of the Hamiltonian. However, the insertion of the sources $O$ can break the unitary invariance giving rise to non-trivial contributions due to the Haar integrations. Haar averages are also central to the study of correlation functions in ensembles of typical states \cite{Pollack:2020gfa, Freivogel:2021ivu}. 

In the following section, we represent the operator $\mathbb{O}$ in a fixed basis, and relative unitary rotations between this fixed basis (in which $\mathbb{O}$ is represented) and the eigenbasis of the Hamiltonian are averaged over using the Haar measure. However, not all observables depend on the Haar average over the unitaries. In \autoref{sec.ope}, we study specific matrix elements of these operators, akin to ETH. In this case the Haar average over the unitary matrices are crucial and play a dominant role in the behaviour of the operators under investigation. In \autoref{sec.g2}, we study the genus-2 spectral form factor. This observable is defined as a trace over the states of the Hilbert space and is insensitive to the Haar average over the unitaries. Both of these observables are important in their own way. The results of \autoref{sec.ope} show that the ORH should be interpreted in a way similar to ETH as discussed in \cite{DAlessio:2015qtq}. The results of \autoref{sec.g2} highlight that the underlying mechanism that determines the genus-2 spectral form factor is the same as the one giving rise to the ORH.

We now note that as before the limit of large Hilbert space dimension $D\gg 1$ allows us to evaluate the integral \rref{eq.SourcedEFT} by saddle point, each saddle giving rise to a perturbative expansion controlled by $D^{-1}=e^{-S}$. Furthermore, the physics of the relevant saddle points is once more controlled by the same causal symmetry breaking structure, so that we are again able to capture the universal chaotic EFT in each sector by a non-linear sigma model with target space $G/K$, as in \rref{eq.SymmetryBreaking} above. Note again that this is true for the simplest of the ten RMT ensembles, namely the unitary class, while the symmetry breaking and thus the sigma-model manifold for the remaining nine take a slightly different form.

\subsection{A tripled Hilbert space}
\label{sec.trip-Hil}
Having provided a pedagogical overview of the tools that we use for our purpose of probing quantum chaos, we move on to develop the formalism that is most useful to the study of CFTs. We start with defining a Hilbert space formed by taking three copies of our original CFT Hilbert space and considering the tensor-product Hilbert space. The advantage of this will become clear in a moment. We refer to it as the tripled Hilbert space,
\begin{equation}
	\mathcal H^{\otimes3} = \mathcal H \otimes \mathcal H \otimes \mathcal H~.
\end{equation}
The Hamiltonian defined on the tripled Hilbert space is,
\begin{equation}\label{eq.tripHam}
	H_{\textrm{full}} = H \otimes \mathds 1 \otimes \mathds 1 + \varkappa_1 \mathds 1 \otimes H \otimes \mathds 1 + \varkappa_2 \mathds 1 \otimes \mathds 1 \otimes H~,
\end{equation}
where, $H$ is the original CFT Hamiltonian. The parameters $\varkappa_i$ are introduced to be able to probe three different moduli in the genus-2 partition function \rref{genus2pf}. In this paper, we are mostly interested in the slice of moduli space where all three moduli are equal and thus often consider $\varkappa_i=1$. In appendix \ref{appA}, we discuss the spectrum of $H_{\rm full}$ and its relationship with the Hamiltonian of the original theory, $H$. We now consider a linear operator acting on this Hilbert space that has the property that one obtains the product of two CFT OPE coefficients when projected onto the energy eigenbasis,
\begin{align} \label{eq.OPEoperator}
	 \langle l \, m \, n| \mathbb O | i \, j \, k\rangle = c^*_{lmn}c_{ijk}  \,.
\end{align}
This should be viewed as a formal definition of the operator, which we can do even without knowing the exact value of the heavy-heavy-heavy OPE coefficients. Nevertheless as we will see, ergodicity and RMT dictates the statistics of the coefficients of this operator $ \mathbb O$. This parallels nicely the situation for ETH, where the microscopic matrix elements are not known but one can derive their statistics from ergodicity, which essentially renders the simple operators Haar-random operators.

Here, and in all subsequent discussions in this paper, we are using the bold notation for traces in the tripled Hilbert space to distinguish it from the trace in the standard Hilbert space.

In fact, we have now translated our CFT question into a usual quantum mechanics setup, so there is no longer really any reason to distinguish our setup with standard questions related to ETH, up to some small differences that we now mention.

The two main differences are related to the structure of the Hilbert space and of the Hamiltonian. Let us focus on a microcanonical window centered at energy $E$.
The first point to note is that the size of the Hilbert space is given by
\be
D^3 \,,\qquad  D=e^{S(E)}=e^{2\pi \sqrt{\frac{c}{3}E}} \,.
\ee
We have used the Cardy formula for the entropy \cite{cardyformula}. It is important \textit{not} to take the energy on the tripled Hilbert space which is $3E$, and think that the dimension of the Hilbert space is then $e^{S(3E)}$. In other words, one should be careful to use the Cardy formula on the original CFT Hilbert space and not on the tripled one.

The next difference is that in usual RMT, we would average over the Hamiltonian on the full Hilbert space. This is a random matrix of dimension $D^3$. Here, we see that in order to connect to a CFT calculation the relevant averaging should be done on a matrix of size $D$, which is then assembled into a bigger matrix following \rref{eq.tripHam}.

Finally, we would like to emphasize again that we have dropped the contribution of descendants, and are only keeping primary operators. One should thus view our setup as a toy-model for CFTs, which captures some but not all of the dynamics. Based on the sparseness of descendants with respect to primaries in CFTs with $d>2$, one expects our toy-model to accurately reflect the behaviour in those CFTs. In $d=2$, the presence of Virasoro symmetry induces extra subtleties, which deserve to be studied in more detail, but at large $c$, primaries should again dominate the dynamics. Moreover, we have not carefully imposed crossing symmetry of all heavy operators.

Keeping these issues in mind, we are now ready to proceed and analyze the consequences of ergodicity for the statistics of heavy operators in a CFT. This can be done thanks to an effective theory that we have reviewed in section \ref{sec.review}.


\subsection{The EFT description of random OPE coefficients}
\label{sec.eft.ope}
Let us now apply the ideas we have reviewed in section \ref{sec.review} to the formalism of section \ref{sec.trip-Hil}. For the operator that encapsulates the OPE coefficients, \eqref{eq.OPEoperator}, we study an observable that we call operator resolvent,
\begin{align}
\label{eq.new-resolvent}
	R(\omega) &= \sum_{i,j} \langle i|\mathbb O | i \rangle \langle j|\mathbb O | j \rangle \delta(E_i-E_j-\omega)\nonumber \\
	&= \int\!\!dE_1dE_2~\rho_{\mathbb O}(E_1) \rho_{\mathbb O}(E_2) \delta(E_1-E_2-\omega)~,\\[10pt]
	\trt [\mathbb O] \trt[\mathbb O] &= \int \!\! d\omega R(\omega)~.
\end{align}
This is a variation of the observable that was studied in \cite{Altland:2021rqn}.
Rewriting the terms containing $\rho_{\mathbb O}(E)$ that appear in the integrand more carefully, we have,
\begin{align}\label{eq.op2pt}
	\trt [\mathbb O] \trt[\mathbb O] &= \int \!\!dE_1\,dE_2 \ \rho_{\mathbb O}(E_1)\rho_{\mathbb O}(E_2) \nonumber \\
						 &= \frac1{2\pi^2}\,\mathfrak{Re}\int\!\!dE_1dE_2\,\Big(\trt\left[G^+\!(E_1)\mathbb O\right] \trt\left[G^-\!(E_2)\mathbb O\right] \nonumber\\
						 &\hspace{4cm}- \trt\left[G^+\!(E_1)\mathbb O\right] \trt\left[G^+\!(E_2)\mathbb O\right]\Big)\nonumber \\
						 &=: \frac1{2\pi^2} \int dEd\omega \, \Big(R^\pm(E,\omega) - R^{\ddag}(E,\omega)\Big)~.
\end{align}
The retarded/advanced Green's functions appearing above are defined by the identity,
\begin{equation}
	G^\pm(E) = \frac1{E\pm i \varepsilon - H_{\rm full}}~.
\end{equation}
The resolvents that are labeled by the causality superscripts are defined by the equivalence of the second and the third line on the RHS of \eqref{eq.op2pt}. These are distinguished from the resolvent refined in \eqref{eq.new-resolvent} by the presence of superscripts as well as different number of arguments. The interesting physics\footnote{More specifically, the $\sigma$-model that governs that late time effective field theory of quantum chaos arises solely from this term. In order to get the full correct answer, including disconnected pieces the other contribution needs of course to be added as well.} arises from the first term, $R^\pm$, in \rref{eq.op2pt} above and we focus our interest on this object from now on. The above quantity is useful to compute a variety of physical observables. The static expression, as it appears in \eqref{eq.op2pt}, computes the variance of the expectation value of the OPE operator. When the integral over the mean energy, $E$, is restricted to a smaller microcanonical energy window, it reproduces the statistics of certain specific OPE coefficients instead. Furthermore, appropriate integral transforms with respect to the energy arguments of the expression gives rise to the genus-2 spectral form factor,
\begin{align}\label{eq.path.resolvents}
	\left|Z[\tau_1,\tau_2,\tau_3]\right|^2 &= \frac1{2\pi^2}\int \!\!dE\, d\omega \, e^{-2\beta E+it\omega} \, \Big(R^\pm(E,\omega) - R^{\ddag}\!(E,\omega)\Big) \\[5pt]
	&\text{where, } \tau_1=\tau_2=\tau_3=t+i\beta \nonumber~.
\end{align}
It now becomes  apparent that the parameters, $\varkappa_i$, are related to the different cycles, $\tau_i$, of the genus-2 surface, see \autoref{fig.g2s}. The observables we are studying here are simply the ratio of determinants as described above \eqref{eq.SourcedEFT},
\begin{align}\label{eq.pathintwrite}
	R^\pm &= -\mathfrak{Re}~ \partial_{h_+}\partial_{h_-} \left. \left\langle \frac{{\rm Det} \left(z^+_1-H_{\rm full} \right) {\rm Det} \left(z^-_2-H_{\rm full} \right)}{{\rm Det} \left(z^+_3-H_{\rm full} +  h_+ \mathbb O \right){\rm Det} \left(z^-_4-H_{\rm full}+ h_- \mathbb O \right)} \right\rangle_{\rm avg.} \right|_{h_\pm=0}
\end{align}
In principle, we have not used any averaging prescription at this point and all the expressions are simply an exact rewriting. It is at this stage, once we represent the observables as determinant operators that we perform some averaging prescription as relevant to the problem.
This ratio of determinants can be expressed in a path integral representation over certain graded fields, $\Psi, \bar\Psi$,
\begin{align}\label{eq.OpDet.path}
	\left\langle \frac{{\rm Det} \left(z^+_1-H_{\rm full} \right) {\rm Det} \left(z^-_2-H_{\rm full} \right)}{{\rm Det} \left(z^+_3-H_{\rm full} +  h_+ \mathbb O \right){\rm Det} \left(z^-_4-H_{\rm full}+ h_- \mathbb O \right)} \right\rangle_{\rm avg.}& \nonumber\\
	&\hspace{-3cm}= \left\langle\! \int \!\! D\bar \Psi D\Psi \ \exp\left[i\bar \Psi \left( \hat z-\!H_{\rm full}\!-\hat h \right) \Psi\right]\right\rangle_{\rm avg.}~.
\end{align}
In the above path integral, we have introduced a component of $\Psi$ fields for each determinant insertion. The determinant insertions in the numerator correspond to fermionic variables while those in the denominator correspond to bosonic variables. Hence the $\Psi$ field has a graded $U(1,1|2)$ structure.\footnote{Understanding the pseudo-Riemannian structure is slightly subtle and arises from having convergent bosonic integrals. See \cite{Altland:2021rqn} for detailed discussion.} Thereby, we have a further enlarged Hilbert space corresponding to the direct product of retarded/advanced (RA), graded boson/fermion (bf) and the tripled Hilbert space. The matrix, $\hat z$, corresponds to energies at which the observables are being probed and $\hat h$ are the source terms,
\begin{eqnarray}
	\hat z &=&\underbrace{ \bigg(E\,\mathds 1 + \left(\frac\omega2 + i \varepsilon \right) \sigma_3\bigg)^{\rm RA}}_{\mathrm{advanced/retarded}} \quad\otimes 
\underbrace{\quad\mathds 1^{\rm bf}\quad}_{\mathrm{boson/fermion}} \quad\otimes  \underbrace{\quad \mathds 1^{\otimes3}\quad}_{\mathrm{tripled}\,\,\mathrm{Hilbert}\,\,\mathrm{space}} ~,\\
	\hat h &=& \begin{pmatrix} h_+&0\\0&0\end{pmatrix} \otimes P_b \otimes \mathbb O +  \begin{pmatrix} 0&0\\0&h_-\end{pmatrix} \otimes P_b \otimes \mathbb O^\dagger~.
\end{eqnarray}
In the first line we have made the tensor product and graded structure of our theory completely explicit. The source terms in the second line are written with this structure in mind. The projector, $P_b$, projects onto the bosonic sector in which the sources are introduced in \eqref{eq.pathintwrite}.

\subsection{Validity of the EFT and Upshot of the Analysis}

To summarize, the EFT provides a way to probe the ergodic limit of chaotic quantum systems. This should also apply to chaotic CFTs: there exists a quantum ergodic regime in which the physical Hamiltonian can be {\it modelled} by an ensemble over Hamiltonians, $H$, chosen from the appropriate Altland-Zirnbauer symmetry class (unitary in the context of the present work). A requirement for the EFT to hold is that it has a finite regime of applicability. In terms of time scales, this is often captured by the Thouless time: the time after which RMT universality kicks in. It is both theory and observable dependent. The main assumption we are making in this paper is that the Thouless time for the thermal trace of the linear operator $\mathbb{O}$ is finite (and parametrically smaller than the Heisenberg time given by the inverse mean-level spacing).  As long as this is true, we can apply our EFT and the framework presented above holds. To really show that the Thouless time is finite, one should carefully integrate out the massive modes as is akin to effective theories and show that the EFT becomes a theory of random Hamiltonians which implements Haar averages. This has been carried out in the SYK model \cite{Altland:2017eao,Altland:2020ccq,Altland:2021rqn}, but a general derivation applicable to arbitrary chaotic systems (including CFTs) is beyond the scope of this work.

We consider two main applications of this EFT. First, we saw that it implements Haar-averages between operator bases and the eigenbases of the Hamiltonian. This procedure should be done within some microcanonical window. In \autoref{sec.ope}, we explicitly preform these Haar averages on $\mathbb{O}$ to study the statistics of OPE coefficients. Since we have a tripled Hilbert space structure, we really have three microcanonical windows which can be taken to be the same or different. Our analysis produces statistics that match the expectations of the OPE Randomness Hypothesis in the ergodic regime. One should remember that while the Haar averages act on the whole microcanonical window, one should only really trust correlations of matrix elements that are close enough in energy space.\footnote{This does not constrain the three microcanonical windows to be close to one another, but simply says that the indices $i,i'$ in $C_{ijk}C^*_{i'j'k'}$ should be close in energy space. The indices $i,j$ are allowed to lie in different microcanonical windows that are very far away in energy space.} As stated above, as long as the Thouless time is finite, the results we present are on solid ground.

The second application is to calculate the genus-2 spectral form factor in the ergodic limit. In \autoref{sec.g2}, we make direct use of the EFT to do so. We find a ramp and a plateau.

\section{Statistics of OPE coefficients}
\label{sec.ope}

As we saw above in section \ref{sec.review}, the EFT of quantum chaos impacts matrix elements of some operator $O$ in a particular way: it essentially implements a Haar average over unitaries between the eigenbasis of the Hamiltonian and that of $O$. Note that this is the logic behind the ETH ansatz in the first place: simple operators are indistinguishable from a typical operator chosen from a Gaussian distribution of operators in a microcanonical window. By performing unitary averages, one can essentially derive the ETH ansatz (see for example \cite{DAlessio:2016rwt}). The role of the EFT is to \textit{derive} that such unitary averages capture the right physics. In this section, we study the implication of these statements for our linear operator $\mathbb{O}$ on the tripled Hilbert space. 

Recall, from \eqref{eq.OPEoperator}, that we can rewrite the product of OPE coefficients as specific elements of the operator $\mathbb O$,\footnote{In a CFT, OPE coefficients have the conjugation property $c_{OOO}=c^*_{O^{\dagger}O^{\dagger}O^{\dagger}}$. This structure must be preserved by the Haar averages, and could influence the RMT universality class of CFTs. We will not be careful about this fact and simply apply unitary averages, but we discuss this point further in the discussion section.  }
\begin{align}\label{eq.csqO}
	c^*_{lmn}c_{ijk} &= \bra{l \, m \, n}  \mathbb O  \ket{i \, j \, k}\,,
\end{align}
where we have lightened the notation a little, denoting energy eigenstates states $\ket{i \, j \, k}$. As we see below, it is important to distinguish whether the triplet of states $\ket{i},\ket{j},\ket{k}$ are in the same microcanonical window or not. We discuss both cases, and start with the simpler case where all three states lie in different microcanonical windows.

\subsection{Different microcanonical ensembles}
To start, note that the EFT of quantum chaos states that the energy eigenstates have a random, uncorrelated projection along some fixed vector(s) in the Hilbert space \cite{mehta-book}. Let us fix some basis for the tripled Hilbert space and the particular choice of basis we make is not important since the energy eigenstates are randomized. We understand the states $\ket{j_{1,2,3}}$ to be vectors in this fixed basis, to distinguish with the eigenbasis of the Hamiltonian which we label by $\ket{i_{1,2,3}}$. In this basis, we can write the operator as
\be
\mathbb{O} =\sum_{j_1,j_2,j_3,j_1',j_2',j_3'} \Omega_{j_1j_2j_3}\Omega_{j_1'j_2'j_3'}^*  \ket{j_1j_2j_3} \bra{j_1'j_2'j_3'} \,,
\ee
We now write the matrix elements of $\mathbb{O}$ in energy eigenstates as

\begin{align}
\label{eq.diff-micro}
	\langle i_i i_2 i_3|\mathbb O |i'_1 i'_2 i'_3\rangle &=\sum_{j_1,j_2,j_3,j_1',j_2',j_3'}\Omega_{j_1j_2j_3}\Omega_{j_1'j_2'j_3'} \langle i_i i_2 i_3 |j_1j_2j_3\rangle \langle j_1'j_2'j_3' |i'_1 i'_2 i'_3\rangle \nonumber\\
	&= \sum_{j_1,j_2,j_3,j_1',j_2',j_3'}\Omega_{j_1j_2j_3}\Omega_{j_1'j_2'j_3'}  U_{i_1j_1}^{(1)} U_{i_2j_2}^{(2)} U_{i_3j_3}^{(3)} U_{j_1' i'_1}^{(1)\dagger} U_{j_2' i'_2}^{(2)\dagger} U_{j_3' i'_3}^{(3)\dagger}
\end{align}
The $U^{(k)}$ are unitaries that implement a change of basis between the fixed basis $\ket{j_k}$ and the eigenstate of the Hamiltonian, $\ket{i_k}$, within a microcanonical window. Note that we have three \textit{different} unitaries $U^{(k)}$, because we are taking the three energy $i_1$, $i_2$, $i_3$ to lie in three different microcanonical windows.
\subsubsection*{Statistical mean of operators}

Ergodicity now implies that the unitaries are random, which means we can integrate over them with the Haar measure. Performing the Haar average over the product of the unitaries gives \cite{Collins2003, collins2006integration},
\begin{align}
\label{eq.signal-diff}
	\overline{\langle i_i i_2 i_3|\mathbb O |i'_1 i'_2 i'_3\rangle} &=\sum_{j_1,j_2,j_3,j_1',j_2',j_3'}\Omega_{j_1j_2j_3}\Omega_{j_1'j_2'j_3'} \int \prod_{k=1}^{3} \left[dU^{(k)}\right] U_{i_1j_1}^{(1)} U_{i_2j_2}^{(2)} U_{i_3j_3}^{(3)} U_{j_1' i'_1}^{(1)\dagger} U_{j_2' i'_2}^{(2)\dagger} U_{j_3' i'_3}^{(3)\dagger} \nonumber \\
									&= \sum_{j_1,j_2,j_3}| \Omega_{j_1,j_2,j_3}|^2 ~~\prod_{m=1}^3 {\rm Wg}(D_i,1,\mathds 1) \delta_{i_mi'_m}
\end{align}
The first argument of the Weingarten functions, Wg, is the dimensionality of the unitary operators. The dimensionality of the unitary operators is the same as that of the microcanonical Hilbert space and are called $D_i$s.
The second argument is $n$, the number of unitary operators inserted in the Haar integral (not counting the daggered-unitary insertions). The third argument is an element of the permutation group, $S_n$. $\mathds 1\in S_n$ corresponds to the identity element that doesn't permute any element in the set. In this case, we simply have
\be
{\rm Wg}(D_i,1,\mathds 1)=\frac{1}{D_i} \,.
\ee
There are several remarkable features that have happened, once the Haar average has been done. First, note that the coefficients are now purely diagonal, the off-diagonal elements having been forced to zero upon doing the Haar average. Second, the only dependence on $\mathbb{O}$ that is left is the term $ \sum_{j_1,j_2,j_3}| \Omega_{j_1,j_2,j_3}|^2$, but this is simply the trace of the operator in the corresponding microcanonical windows.\footnote{We have a microcanonical trace because the indices, $j_1,j_2,j_3$, only take values corresponding to the states that lie in the relevant microcanonical window.} All dependence on the individual $\Omega_{j_1,j_2,j_3}$ or on the choice of basis $\ket{j_{1,2,3}}$ has been washed out. Since the trace is basis independent, it is most convenient to write it in the energy basis and we thus have that the coefficient of the Kronecker delta is
\be
\frac{1}{D_1 D_2 D_3 }\trt \mathbb{O} = \frac{1}{D_1 D_2 D_3 }\sum_{ijk} |c_{ijk}|^2 \,,
\ee
which is the averaged sum over OPE coefficients squared. This is \textit{coarse-grained} information which we have access to without knowing the detailed microscopics. In ETH, this would correspond to the diagonal smooth function $f(\bar{E})$. 
Just like in ETH, this function of the mean energies is not fixed from first principles, and can be theory dependent. For CFTs in general dimensions, it is currently unknown how to constrain the heavy-heavy-heavy OPE coefficients (this is explored in \cite{6ptasymptotics}). However, for $d=2$, Virasoro symmetry fixes the averaged OPE coefficients. This is a result of modular invariance of the genus-2 partition function. Explicit expressions for the averaged OPE coefficients can be found in \cite{Cardy:2017qhl,Collier:2019weq}.

It is worth mentioning that the OPE coefficients are invariant under the permutation of the three indices, and so the statistics should be built in a way that preserves this symmetry. This is a minor technical step that we have not taken into account in the presentation so far, for the sake of clarity. It would be straightforward to do so, but we leave this procedure as being done implicitly to avoid cluttering expressions.
\subsubsection*{Statistical variance  of operators}
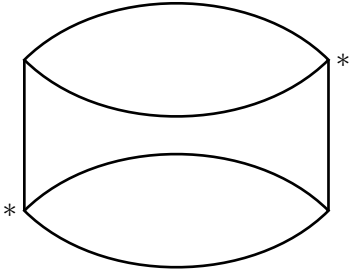
\begin{figure}[t]
\centering
\begin{tikzpicture}
	\draw[line width = 1pt] (0,0) node(st1) {$*\quad$}  ..controls (1,1) and (3,1) .. (4,0) ..controls (3,-1) and (1,-1) .. cycle;
	\draw[line width = 1pt] (0,2)  ..controls (1,3) and (3,3) .. (4,2) node(st2) {$\quad*$}  ..controls (3,1) and (1,1) .. cycle;
	\draw[line width = 1pt] (0,0) -- (0,2);
	\draw[line width = 1pt] (4,0) -- (4,2);
\end{tikzpicture}
\caption{Sunset diagrams are the diagrammatic representation of the contractions depicted in \eqref{eq.sunset}. Each vertex of the diagram represents an insertion of a OPE coefficient. The ones corresponding to the insertions of $c^*$s are marked with a star, while the unmarked vertices are the $c$ insertions. Each edge in the diagram corresponds to the index contractions between OPE coefficients.}
\label{fig.sunset}
\end{figure}

Having obtained the mean of the matrix elements of $\mathbb{O}$, we now proceed to study their statistical variance. We thus wish to compute
\begin{equation}
	\overline{\langle i_i i_2 i_3|\mathbb O |i'_1 i'_2 i'_3\rangle^2}- \overline{\langle i_i i_2 i_3|\mathbb O |i'_1 i'_2 i'_3\rangle}^2~,
\end{equation}
which can be done in a similar fashion by an appropriate integration over the unitary matrices using the Haar measure. We give the explicit and detailed computation in appendix \ref{app.variance}. The leading order contribution to the variance is given by 
\bea\label{eq.variance}
	\overline{\langle i_i i_2 i_3|\mathbb O |i'_1 i'_2 i'_3\rangle^2}- \overline{\langle i_i i_2 i_3|\mathbb O |i'_1 i'_2 i'_3\rangle}^2 ~ &\approx& ~ \left(\frac{1}{D_1 D_2 D_3}\right)^2 \Big[\ ~\trt \mathbb{O} ~ \trt \mathbb{O}  \\
	&\hspace{-5.75cm}+&\hspace{-3cm} \trt_{\rm sunset} \mathbb{O}^2\Big(\delta_{i_1i'_1}\delta_{i_2i'_2} + \delta_{i_1i'_1}\delta_{i_3i'_3} + \delta_{i_3i'_3}\delta_{i_2i'_2} ~+ \delta_{i_1i'_1} + \delta_{i_2i'_2} + \delta_{i_3i'_3}\Big)\Big]~ \,, \notag
\eea
where we have defined the sunset trace contraction as  
\be\label{eq.sunset}
\trt_{\rm sunset} \mathbb{O}^2=c_{j_1j_2j_3}c^*_{j_1j_2k_3} ~ c_{k_1k_2k_3}c^*_{k_1k_2j_3} \,.
\ee 
A diagram of this contraction is depicted in \autoref{fig.sunset}.
This particular sum of OPE coefficients appears in the sunset channel decomposition of a genus-3 partition function, hence the name we have chosen for the trace.

It is important to note that the mean is as large as the variance (and for that matter all the subsequent moments of the operator $\mathbb O$). Hence, we can not apply the central limit theorem and write the operator $\mathbb O$ using random variables which are approximately Gaussian. If we wish to write an ansatz for $\mathbb O$ which resembles ETH, we can write
\begin{align} \label{resultdiffwindow2}
	c^*_{i_1i_2i_3} c_{i'_1i'_2i'_3}  \approx  \frac{\trt \mathbb{O}}{D_1 D_2 D_3} (\delta_{i_1,i_1'}\delta_{i_2,i_2'}\delta_{i_3,i_3'}+R_{i_1,i_2,i_3,i_1',i_2',i_3'})  \nonumber \\ + \frac{\sqrt{ \trt_{\rm sunset} \mathbb{O}^2}}{D_1 D_2 D_3}\textrm{Sym} &\left[\delta_{i_1,i_1'} R_{i_2,i_3,i_2',i_3'}+\delta_{i_1,i_1'}\delta_{i_2,i_2'} R_{i_3,i_3'} \right] \,,
\end{align}
where the $\textrm{Sym}$ stands for $S_3$ symmetrizing between the indices 1, 2 and
3, and simultaneously acting on 1', 2' and 3' in the same way. We must however be very careful with what we mean by such an equation. 

Contrary to ETH, the variables $R$ above are \textit{not} approximately Gaussian random variables. They are statistical variables that have zero mean and unit variance, but they have non-trivial higher moments that are of the same size as the first and second moments. The reason for this is that we are trying to write random variables directly for the operator $\mathbb O$, which is really a composite tensor made out of two $c_{ijk}$. It is the $c_{ijk}$ that are approximately Gaussian random, and the non-trivial random matrix in the diagonal $R_{i_1,i_2,i_3,i_1',i_2',i_3'}$, which is the same size as the mean, comes from the fact that there is a cross Wick contraction in the variance of $\mathbb O$ which involves four $c_{ijk}$. One can make an intuitive parallel with large $N$ factorization in large $N$ gauges theories. At infinite $N$, the theory is a generalized free theory and all correlation functions are obtained from Wick contractions. However, connected correlation function of multi-trace operators are non-trivial. This does not mean that the underlying theory is not Gaussian, a free theory is obviously Gaussian, but simply that we are considering composite operators. The same logic applies to \rref{resultdiffwindow2}.

To sum up, it is worth emphasizing what \rref{resultdiffwindow2} really means. The microscopic definition of the operator $\mathbb{O}$ was given in \rref{eq.csqO} and seems in tension with the result \eqref{resultdiffwindow2}. One should view these results in the same way as one views ETH: there is a microscopic definition of the matrix elements, and there is the ETH ansatz which is true in a statistical sense. It should be understood as statistically true, once sampled over enough individual coefficients. While the general structure of our result follows closely that of matrix elements in the ETH, there are also important distinctions to take into account, in particular the composite nature of the operator $\mathbb O$. We discuss them in section \ref{ETHcomp}, but we first give the general expression when all microcanonical windows are the same.


\subsection{A single microcanonical ensemble \label{singlewindsec} }
In this section we analyse the behaviour of the operator elements when the external states belong to the same microcanonical window of energies. In this case, all unitaries that enter in \rref{eq.signal-diff} are the same. Once integrated using the Haar measure, they produce new structure because more Wick contractions are allowed. Let us start once more by calculating the mean value of operators.
\subsubsection*{Statistical mean of operators}
We have
\begin{align}
\label{eq.same-micro}
	\langle i_i i_2 i_3|\mathbb O |i'_1 i'_2 i'_3\rangle &=\sum_{j_1,j_2,j_3,j_1',j_2',j_3'}\Omega_{j_1j_2j_3}\Omega_{j_1'j_2'j_3'}  U_{i_1j_1} U_{i_2j_2} U_{i_3j_3} U_{j_1' i'_1}^{\dagger} U_{j_2' i'_2}^{\dagger} U_{j_3' i'_3}^{\dagger}
\end{align}
The Haar integration over $U$ is more complicated but the final result is
\begin{align}
\label{eq.signal-same}
	\overline{\langle i_i i_2 i_3|\mathbb O |i'_1 i'_2 i'_3\rangle} &= \trt \mathbb{O}  ~ \Big( {\rm Wg}(D_1,3,\mathds 1) + 3{\rm Wg}(D_1,3,2\cdot1) + 2{\rm Wg}(D_1,3, 3) \Big) \nonumber \\
	& \qquad \bigg(\delta_{EE'} + \delta_{i_1i'_2}\delta_{i_2i'_3}\delta_{i_3i'_1} + \delta_{i_1i'_3}\delta_{i_2i'_1}\delta_{i_3i'_2} \nonumber \\
	&\qquad \qquad + \delta_{i_1i'_1}\delta_{i_3i'_2}\delta_{i_2i'_3} + \delta_{i_1i'_3}\delta_{i_2i'_2}\delta_{i_3i'_1} + \delta_{i_1i'_2}\delta_{i_2i'_1}\delta_{i_3i'_3} \bigg) \nonumber \\
	&= \trt \mathbb{O}  ~ \frac1{D_1(D_1+1)(D_1+2)} ~ [\delta_{EE'}]_{S_3} ~.
\end{align}
Note that in this case we have both diagonal as well as ``off-diagonal" terms. However, the off-diagonal terms are essentially the permutations of three indices labelling the OPE coefficients, so they are rather trivial. To avoid clutter, we denote the $S_3$ symmetrised terms in the primed coordinates by a subscript, $[\delta_{EE'}]_{S_3}$.

To leading order in the large $D_1$ limit, the coefficient of the delta function is simply $\trt\mathbb{O}/D_1^3$ which agrees with \eqref{eq.signal-diff}, once we have set the three windows to be the same size. This makes sense, we wouldn't expect there to be a big different between the same and distinct microcanonical windows. There are only two small differences: first, we see that the indices are symmetrized, which is natural in this case. Second, we see that there are further exponentially suppressed corrections to the leading coefficient.

\subsubsection*{Statistical variance of operators}
Once again, to be able to describe the random nature of the matrix elements of $\mathbb O$, we need to compute their variance. In the present scenario, we find
\begin{align}\label{eq.variance.same}
	\overline{\langle i_i i_2 i_3|\mathbb O |i'_1 i'_2 i'_3\rangle^2}- \overline{\langle i_i i_2 i_3|\mathbb O |i'_1 i'_2 i'_3\rangle}^2&\nonumber\\
																	    &\hspace{-3.25cm}\approx {\rm Wg}(D_1,6,\mathds 1)\bigg[ ~ \trt\mathbb O \trt\mathbb O \left( 1+\delta_{i_1i_2} + \delta_{i_2i_3} + \delta_{i_3i_1} + \delta_{i_1i_2} \delta_{i_2i_3} \delta_{i_3i_1} \right) \times (\cdots)'  \nonumber \\
									  &\hspace{-0.5cm}+ ~ \trt_{\rm sunset}[\mathbb O^2] \Big(\delta_{i_1i'_1}\delta_{i_2i'_2} + \cdots 647 \text{ terms } \cdots \Big) \bigg]~.
\end{align}
The term $(\cdots)'$ represents the same parenthesis that precedes it, but with primed indices. The 647 other terms correspond to all possible combinations of delta functions and symmetrizations thereof. The gory details are given in appendix \ref{haarsame}.

Therefore the statistical properties of the heavy-heavy-heavy OPE coefficients as encapsulated by the operator, $\mathbb O$, can be effectively given written as,
\begin{align} \label{Oonewind}
	c^*_{i_1i_2i_3} c_{i'_1i'_2i'_3}   &\approx  \frac{\trt \mathbb{O}}{D_1^3} (\textrm{Sym}_{S_3}[\delta_{i_1,i_1'}\delta_{i_2,i_2'}\delta_{i_3,i_3'}]+R^s_{i_1,i_2,i_3,i_1',i_2',i_3'} \nonumber \\
	&\quad +\frac{\sqrt{ \trt_{\rm sunset} \mathbb{O}^2}}{D_1^3}\textrm{Sym}_{\textrm{full}} \left[\delta_{i_1,i_1'} R_{i_2,i_3,i_2',i_3'} +\delta_{i_1,i_1'}\delta_{i_2,i_2'} R_{i_3,i_3'} \right] \,,
\end{align}
where $R^s$ is now a symmetric random tensor and $\textrm{Sym}_{\textrm{full}}$ is now fully symmetrizing in all indices using the symmetric group $S_6$. Once again, one must be careful with the interpretation of this formula, the statistical variables $R$ are not approximately Gaussian random, but their mean vanishes and they have unit variance.
We see that this is a more sophisticated symmetrized version of \rref{resultdiffwindow2}, but other than that it shares the same general structure. This gives us a measure of the variance of the OPE coefficients as conjectured in \cite{Belin:2020hea}. We now move to discussing the difference and similarities between $\mathbb{O}$ and an operator in ETH.

\subsection{Comparison with conventional ETH \label{ETHcomp}}
The framework we have developed to study the heavy-heavy-heavy OPE coefficients naturally shares many features with conventional ETH where the objects of study are the matrix elements of operators. This follows from the fact that we have defined a linear operator on a tripled Hilbert space. Nevertheless, there are also a few differences that we would like to underline.

The first difference we have come across is the fact that the diagonal and off-diagonal matrix elements of $\mathbb{O}$ are of the same size. This should be contrasted with ETH, where the diagonal contribution given by the microcanonical one-point function is expected to be much larger.\footnote{Note that this separation between diagonal and off-diagonal parts is not universal and in particular in 2d CFTs, the diagonal term is kinematically exponentially suppressed because one-point functions of primary operators vanish on the cylinder.} The reason for this is that the operator $\mathbb{O}$ is not the most general operator that acts on a Hilbert space of size $D^3$,\footnote{In fact, it is a projector. Had we picked the fixed basis to be the eigenbasis of $\mathbb{O}$, we would have had a single non-zero eigenvalue, which shows the operator is very constrained.} but rather is a composite operator made out of two tensors $c_{ijk}$. It is thus specified by $D^3$ numbers, rather than a generic operator which would require $D^6$ terms. Comparing to ETH, this would be similar to defining a vector $v$ such that
\be
R_{mn}=v_m v_n^* \,,
\ee
Clearly, the most general matrix is not of this type. Now, if $v$ effectively becomes averaged over, the diagonal and off-diagonal elements are on different footing since the diagonal elements are real and positive numbers. Something similar is happening for us with the operator $\mathbb{O}$.

The fact that the off-diagonal matrix elements are of the same size as the diagonal ones should however be seen as a feature, not a bug. The ergodic nature of the OPE coefficients is consistent with the observed behaviour of the operator $\mathbb O$, that can be regarded as a composite operator in terms of the OPE coefficients $c_{ijk}$. The Gaussian nature of the $c_{ijk}$ imply ``large" higher moments for  $\mathbb{O}$. This is why the variance was the same order as the mean, because there is a non-trivial Wick contraction in the variance of $\mathbb{O}$, which contains four $c_{ijk}$. The same is true for higher moments, that all have the same size as the mean. We would like to emphasize that this is not inconsistent with the $c_{ijk}$ being Gaussian, but rather happens precisely because they are!

A second difference with ETH comes from the size of the true non-Gaussianities, which in our case are measured by $\sqrt{ \trt_{\rm sunset} \mathbb{O}^2}$. In ETH, one finds that the variance is suppressed by $e^{-S/2}$ as long as the microcanonical two-point function is of the same order as the microcanonical one-point function. Here, the condition for the true non-Gaussianities to be small is that 
\be
\sqrt{ \trt_{\rm sunset} \mathbb{O}^2} \ll \trt \mathbb{O} \,.
\ee
This can be viewed as the condition for the OPE coefficients to be approximately Gaussian random variables. It is currently unknown if this is true in all CFTs, but it has been proven in $d=2$ \cite{Belin:2021ryy}. It is however important to underline that it is \textit{not} an output of our analysis, but rather has to be assumed as an extra condition.

Finally, we would like to emphasize that we did not do a $D^3$ unitary average over the basis of the Hamiltonian, which is what would be required if we had naively applied standard ETH to $\mathbb O$ with a random Hamiltonian of size $D^3$. If we would have done this larger unitary average, $\mathbb O$ would have behavd as
\begin{align}
	\langle E|\mathbb O| E'\rangle \approx \frac{\trt \mathbb O}{D^3} \Big(\delta_{EE'} + R_{EE'}\Big)~,
\end{align}
which misses the contributions of the sunset diagrams. As we now see, observables like the genus-2 spectral form factor do not receive contributions from the Haar-average over the unitary operators, and thus do not actually perceive the triple tensor-product structure of the Hilbert space. We discuss this in further in the following section, as we now turn to the study of the genus-2 spectral form factor using the EFT of quantum chaos.

\section{EFT for the Genus-2 Spectral Form Factor}
\label{sec.g2}
In this section we change gears and compute the genus-2 spectral form factor, \eqref{genus2sffintro},
\begin{align} \label{genus2sffintro.sec4}
	F(t_1,t_2,t_3) &= |Z(\tau_1+it_1,\tau_2+it_2,\tau_3+it_3) |^2 \,,\\
	&Z(\tau_1,\tau_2,\tau_3) = \sum_{O_1,O_2,O_3} |C_{123}|^2 q_1^{\Delta_1} q_2^{\Delta_1} q_3^{\Delta_3} \,, \qquad q_i=e^{2\pi i \tau_i} \,.
\label{eq.g2.sum.explct}
\end{align}
As defined, the genus-2 spectral form factor involves a sum over all the operators of the CFT. Because of the state-operator correspondence, this amounts to a sum over the entire Hilbert space of the theory. Nevertheless, at sufficiently high ``temperatures" (i.e. for small moduli) the dominant contribution arises from the high energy part of the spectrum. For holographic CFTs, as long as one is above the Hawking-Page temperature, this would probe the chaotic sector of the spectrum. Alternatively, one could restrict the sum in \eqref{eq.g2.sum.explct} to a microcanonical window within this high energy spectrum. This microcanonical genus-2 spectral form factor would capture the chaotic physics that we are interested in  equally well. For simplicity, we consider the sum over the entire Hilbert space of our theory, while pointing out the differences with a microcanonical version of the spectral form factor when relevant.\footnote{Note that we are considering CFTs, so formally the dimension of the full Hilbert space $D=\infty$. In the end, we study a fixed microcanonical window of size $D_1$ which in practice corresponds to setting $D=D_1$. We keep $D$ general is what follows, see the discussion after \rref{eq.HHA}.}

In section \ref{sec.eft.ope}, we discussed how one can write the genus-2 spectral form factor as a correlation function of a linear operator that acts on the tripled Hilbert space. This facilitates a rewriting of the genus-2 partition function as the path integral
\begin{align}\label{eq.resolvent.path}
	R^\pm &= - \mathfrak{Re} \left[\partial_{h_+}\partial_{h_-} Z[\hat h]\right], \nonumber \\
		Z[\hat h] &= \left\langle \int \!\! D\bar \Psi D\Psi \ \exp\left[i\bar \Psi \cdot \left( \hat z-H_{\rm full}-\hat h \right)\cdot \Psi\right] \right\rangle_{\rm avg.}~.
\end{align}
An important difference from the observables studied in the preceding sections is that the genus-2 spectral form factor involves traces over either the entire Hilbert space or a part of it - an appropriate microcanonical window. The cyclicity of traces implies that the unitaries that rotate the physical states of the Hilbert space cancel against their complex conjugates. Consequently, the genus-2 spectral form factor only depends on the statistics of the eigenvalues in the ergodic limit but not on that of the eigenstates. As we shall see below, this implies that in the quantum ergodic limit, the information about the product nature of the underlying Hilbert space is not relevant. Although, when working with microcanonical windows one would need to restrict the auxilliary graded vectors introduced in the above equation, as well as the corresponding unitary rotations to only the relevant sector of the Hilbert space.

We now bring the formalism described in section \ref{sec.review} to bear on the genus-2 spectral form factor. This involves writing down the effective field theory of chaos based on the tripled Hilbert space we made copious use of in this work. The EFT of quantum chaos is a theory on the space of Hamiltonians, whose dynamics captures the approach of a generic chaotic Hamiltonian to the predictions of random-matrix theory in the ergodic limit (see e.g. \cite{Altland:2020ccq,Altland:2021rqn}). 

In order to proceed, let us write the Hamiltonian, $H$, in a complete basis of hermitian matrices, $X_a$, in $D$-dimensions, which we then adapt to the tripled Hilbert space

\begin{align}
	\mathbb X_a &\equiv X_a\otimes \mathds 1 \otimes \mathds 1 + \varkappa_1 \mathds 1 \otimes X_a \otimes \mathds 1 + \varkappa_2 \mathds 1 \otimes \mathds 1 \otimes X_a,
\end{align}
This basis is labelled by the parameter $a$ which takes values,  $a = 0,1,2,\ldots , D^2-1$ and $X_0 = \mathds 1$.  We can then expand the tripled Hamiltonian in this basis as\footnote{In appendix \ref{appA}, we discuss the spectrum of $H_{\textrm{full}}$ and its relationship with the Hamiltonian of the original theory, H.}
\begin{align}
	H_{\rm full} &= a^a \mathbb X_a 
\end{align}

The ergodic limit is characterised by the indistinguishability of the physical Hamiltonian from a typical representative of the corresponding RMT ensemble. How a given chaotic Hamiltonian approaches this limit is a highly interesting subject,  \cite{Altland:2020ccq,Altland:2021rqn}, but for the present purposes we may simply assume that the theory has such an ergodic limit. The strength of our formalism is that it is well-suited to study the non-universal modes that lead to deviation from the ergodic limit at times earlier than the {\it Thouless/ergodic time}. These non-universal modes, referred to as massive modes, arise from the fact that the matrix elements of a physical Hamiltonian are highly correlated. Outside the ergodic limit, this is reflected in correlations between the $a^a$ parameters. However, in the quantum ergodic limit, random-matrix universality implies that the individual elements of the Hamiltonian, $H_{ij}$, are chosen from a Gaussian distribution. This translates to performing a Gaussian disorder average over the $a^a$ parameters with a probability distribution given by,
\begin{equation}
	P(a^a) = \exp\left[-\frac1{2\bar a} \left(a^a\right)^2\right]~,
\end{equation}
for some fixed $\bar a$. Performing the integral over these parameters gives us the expression
\begin{equation}
	Z[\hat h] = \!\! \int \!\!\!D\bar \Psi D\Psi \, \exp\left[i\bar \Psi \cdot \left( \hat z-\hat h \right)\cdot \Psi - \frac {\bar a}2 \sum_a\left(\bar \Psi \mathbb X_a \Psi \right)^2\right]~.
\end{equation}
whose last term is quartic in the graded fields $\Psi$. This  quartic term can be treated by a Hubbard-Stratanovich transformation to rewrite the path integral in terms of the new fields, $A^a$,
\begin{align}
	Z[\hat h] &=  \int DA^a D\bar \Psi D\Psi \ \exp\left[i\bar \Psi \cdot \left( \hat z-\hat h \right)\cdot \Psi +i \sum_a \strt\left[A^a \left(\mathbb X_a \Psi \bar \Psi \right)\right] \right. \nonumber \\
	&\hspace{5cm} \left. - \frac 1{2\bar a} \sum_a \strt \left[ (A^a)^2\right]\right]\,.
\end{align}
The resulting action is quadratic in the $\Psi$ fields and which can therefore be integrated exactly, so that finally we arrive at the expression
\begin{align}
		 Z[\hat h] &=  \int DA^a \exp\left[ -\frac1{2 \bar a}\sum_a\strt\left[(A^a)^2\right] - \strt\ln\left[ \hat z - \hat h + A^a\mathbb X_a \right] \right]\,.
\end{align}
In the above and in some of the following expressions we are not taking care of the normalisation factors which can be easily fixed by requiring that the final expression for the path integral should be equal to 1 when all the sources are switched off. This is true for the simple reason that the determinants that appear in the numerator and the denominator of \eqref{eq.pathintwrite} are identical in the absence of sources.

The final expression above is amenable to saddle point analysis when the size of the Hilbert space is large. The saddle point equations are
\begin{equation}\label{eq.full.saddle}
	A^a = - \bar a \left( \hat z + A^a\mathbb X_a \right)^{-1}\, \mathbb X_a~.
\end{equation}
Note that we have dropped the source term in the above saddle equation. Following a standard treatment, the sources are treated perturbatively and the operators are assumed to not backreact on the solutions of the saddle equations.\footnote{This is an assumption. We comment on this point further in the discussion section.} At this stage, we consider the homogeneous Hilbert space ansatz according to which the saddle point solution of the above path integral is homogeneous in the complete $D^3$ dimensional Hilbert space,
\begin{equation}\label{eq.HHA}
	A^a = \hat y \,\delta^a_0 \, \mathds 1^{\otimes 3}~.
\end{equation}
The physical motivation and the validity of this ansatz is discussed in great detail in \cite{Altland:2020ccq}. When working with a microcanonical window instead of the entire Hilbert space, the homogeneous Hilbert space ansatz is restricted only to the relevant part of the Hilbert space in this equation. This implies that the identity matrix, $\mathds 1^{\otimes 3}$, is not an identity matrix in the entire Hilbert space but only a subsector of it. The factors of Hilbert space dimensions that appear in the following discussion will therefore get modified accordingly to reflect the dimensionality of the microcanonical Hilbert space (effectively, this means taking $D=D_1$ from now on).

The path integral simplifies on taking the ansatz, \eqref{eq.HHA}, also making the validity of the saddle point approximation more apparent in the large $D$ limit,
\begin{align}\label{eq.homogeneous-action}
	Z[\hat h] & =\int \!\!D\hat y \exp\Bigg[ \!\!-\frac{D^3}{2 \bar a}\sum_a\strt\left[(\hat y)^2\right] - D^3 \, \strt\ln\left[ \hat z + \gamma \hat y  \right] -\strt\ln\left[\mathds 1 + \left(\hat z + \gamma \hat y\right)^{-1} \hat h\right]\!\!\Bigg]~,\\
	&\hspace{5cm}\gamma := 1+\varkappa_1+\varkappa_2~.
\end{align}
Note that we have chosen to write the term containing the sources seprately. Because we evaluate the path integral at $\hat h=0$, this term is expanded in powers of $\hat h$ starting at linear order. In the $D\gg1$ limit, the solutions of the saddle point equation, \eqref{eq.full.saddle}, simplifies to take the form
\begin{align}
	\hat y &= - \bar a \left( \hat z + \hat y \right)^{-1}~,\\
	\Rightarrow \hat y &= -\frac E{2 \gamma} + i \Lambda \sqrt{\bar a - \frac{E^2}{4\gamma^2}},
\end{align}
where the matrix $\Lambda$  determines the different choices of saddle, as
\begin{align}
	\Lambda_0 &= \sigma_3^{\rm RA} \otimes \mathds 1^{\rm bf} \label{eq.leading-saddle}\\
	\Lambda_{\rm AA} &= \sigma_3^{\rm RA} \otimes \sigma_3^{\rm bf} \label{eq.subleading-saddle}
\end{align}
Following standard convention, we refer to these as the \emph{standard saddle} and  the \emph{Andreev-Altschuler saddle}, respectively. The standard saddle point corresponds to the following density of states in the tripled Hilbert space,
\begin{equation}\label{eq.TripledDensityOfStates}
	\rho(E) = \frac{D^3}{\pi \sqrt{\bar a} \gamma} \sqrt{1 - \frac {E^2}{4 \bar a \gamma^2}}, \quad \Delta(E) = \frac1{\rho(E)}~.
\end{equation}
We have also defined the mean level spacing, $\Delta(E)$ as the inverse of the density of states, $\rho(E)$.  One way of seeing \rref{eq.TripledDensityOfStates} is to use \rref{eq.OperatorWeightedDensity} - \rref{eq.SourcedEFT} to compute
\begin{equation}\label{eq.1ptdense}
\rho(E) :=  \rho_{ \mathds 1}(E)  = \frac{1}{\pi} {\rm Im}\,\partial_{h_1} Z[\hat h] \Bigr|_{\hat h =0}\,.
\end{equation}
Notice that as long as the energy arguments that appear in $\hat z$ are all the same, the path integral \eqref{eq.homogeneous-action} is symmetric under the $U(1,1|2)$ rotations in the graded spaces,
\begin{equation}
	\Psi \to g \cdot \Psi , \quad g\in U(1,1|2)~.
\end{equation}
which translates to following transformation of the Hubbard-Stratonovic field, $\hat y$,
\begin{equation}
	\hat y \to g \cdot \hat y \cdot g^\dagger~.
\end{equation}
Both the saddle points break this $U(1,1|2)$ symmetry spontaneously to $U(1|1)\times U(1|1)$. In fact, when $\omega=0$, the entire coset manifold, $U(1,1|2)/U(1|1)\times U(1|1)$, is a space of solutions of the saddle point solutions. We parametrise this manifold as a `rotation' of the standard saddle point,
\begin{align}\label{eq.pions2}
	T \Lambda_0 T^{-1} \in \frac{U(1,1|2)}{U(1|1)\times U(1|1)}.
\end{align}
The Andreev-Altschuler saddle point lies on this manifold and corresponds to,
\begin{equation}
	T = T_0 = \mathds 1^{RA} \otimes P_b + \sigma_1^{RA} \otimes P_f~.
\end{equation}
The fact that saddle point solutions break the symmetry spontaneously can be argued as follows.
The sign of the regulator, $\varepsilon$, that differentiates between the advanced and the retarded sector leads to this spontaneous symmetry breaking. In a fashion the reader might be familiar from the treatment of advanced and retarded correlation functions in elementary QFT, the regulator displaces the poles in the integration plane of the eigenvalues of $\hat y$ by a small amount along the imaginary direction. The integration contour runs along the real axis for the bosonic components of the graded matrix, $\hat y$. The presence of the infinitesimally displaced pole restricts the deformation of this integration contour to only one of the two choices for the saddle point solutions. This choice is different for different causalities, hence leading to the spontaneous symmetry breaking. The fermionic sector has no such restrictions because of the lack of poles in the integration plane. For a more detailed technical discussion, we refer the reader to \cite{Altland:2021rqn, Haake-book}.

The precise nature of the structure of symmetry breaking relies on the observable under study. For example, in the computation of the density of states in \eqref{eq.1ptdense}, $\rho(E)$ involves only one of the two causality sectors. Therefore the symmetry breaking is irrelevant in this case. Similarly, for the computation of higher point functions, we need to introduce additional copies of the Hilbert space corresponding to additional insertion of the determinant operators. In this case too, the symmetry breaking will be different. Such observables are not the subject of current work.

The $U(1,1|2)$ symmetry is explicitly broken when $\omega\neq0$, which corresponds to different values of $z_i$s. In this case, we do not have a full coset manifold worth of saddle points solutions. However, the standard saddle and the Andreev-Altschuler saddle continue to be solutions to the saddle point equations. The additional action-cost associated with the points of the coset manifold can be computed in the limit $s = \pi \omega/\Delta \ll 1$. We treat $s = \pi \omega/\Delta$ as a perturbative parameter and expand the action, \eqref{eq.homogeneous-action}, as a power series in this parameter. Under such a treatment, various points on the coset-manifold can therefore be regarded as \emph{pseudo-Goldstone bosons}.

It is useful to parametrise the points on the coset manifold in terms of \emph{pion fields}, $W$,
\begin{align}\label{eq.pions}
	Q = T \Lambda_0 T^{-1} \in \frac{U(1,1|2)}{U(1|1)\times U(1|1)}, \quad T = e^{-W}~,\quad
	W = -\begin{pmatrix}0&B\\ \tilde B&0\end{pmatrix}~.
\end{align}
On substituting the above parametrisation of the coset manifold into the action, \eqref{eq.homogeneous-action}, and performing the perturbative expansion in small $\omega$ one obtains the leading order term,
\begin{equation}\label{eq.quark-mass}
	2i \pi \frac\omega{ \Delta(E) } \ \strt\left[B \tilde B\right]~.
\end{equation}
The above term can be compared to the ``quark mass'' for the pions in the chiral perturbation theory. Following a top down approach, for complex Hamiltonians one can write down an effective action on the coset manifold in increasing orders of $\omega$ as well as $B,\tilde B$,
\begin{align}
	\int \!\!dQ\, e^{-S[Q;\omega]}, \qquad Q \in \frac{U(1,1|2)}{U(1|1)\times U(1|1)}~,
\end{align}
in the same spirit as the chiral perturbation theory.
The integration on the coset manifold can be performed exactly \cite{Haake-book}. However, because the integration is one-loop exact, \cite{Haake-book, Efetov-book}, it is sufficient to perform the integration perturbatively in the pion fields around the individual saddle points. It is for this reason we have expanded the action in $B,\tilde B$ fields and kept only the leading quadratic term in above equation.

To compute the resolvent, \eqref{eq.resolvent.path}, we need to take the derivative of the path integral, \eqref{eq.homogeneous-action}, with respect to the sources, $h_\pm$. This amounts to expanding the last term in this action that depends on the source to quadratic order in $\hat h$. Taking derivatives of this term with respect to the sources introduces certain pre-exponential terms in the $\sigma$-model path integral,
\begin{equation}\label{eq.pre-exp}\begin{aligned}
	\left( \frac1{\bar a\gamma^2} + \frac{4\pi^2 \rho^2(E)}{D^6} \strt\left[ B \tilde B P_b\right] \strt\left[ \tilde B B P_b\right]\right) \trt\mathbb O \, \trt \mathbb O^\dagger\\
	- \frac{4\pi^2 \rho^2(E)}{D^6} \strt\left[ B P_b \tilde B P_b \right] \trt[\mathbb O \mathbb O^\dagger] + \cdots~.
\end{aligned}\end{equation}
Here, we have explicitly written only the terms that involve non-trivial contraction of $B, \tilde B$ fields. There are other terms that evaluate to zero and aren't shown above. From \eqref{eq.quark-mass} one can deduce the following contraction rules between the pion fields:
\begin{align}
\label{eq.BBtprop1}
	\strt[{B}\cdot X] \, \strt[ { \tilde B} \cdot Y] &= \frac1{2\pi \rho(E)} \frac i{\omega+i0^+} \strt[X Y]~,\\
\label{eq.BBtprop2}
	\strt[\source{B} \cdot X \cdot \target{\tilde B}\cdot Y] &= \frac1{2\pi \rho(E)} \frac i{\omega+i0^+} \strt[X] \strt[ Y]~.
\end{align}
Using these contraction rules, one can solve \eqref{eq.pre-exp} to get the following answer for the resolvent,
\begin{equation}\label{eq.std.saddle.ans}
	R(\omega) = \left(\frac1{\bar a \gamma^2} - \frac1{D^6 \omega^2}\right) \tr \mathbb O \, \tr \mathbb O^\dagger~.
\end{equation}
Note that contractions between the $B, \tilde B$ that have a common index is zero. For example, one does not receive a contribution from contracting the fields within the same $\strt$ terms. This is because it corresponds to $X=\mathds 1$ in \eqref{eq.BBtprop2} and $\strt \mathds 1=0$.
As we remarked above, for small $s$ when the symmetry breaking is small, one needs to perform the integration over the entire coset manifold. However, one-loop exactness of the path integral implies that the exact answer is the sum over one-loop contributions around the individual saddle points. The result in \eqref{eq.std.saddle.ans} is the contribution around the standard saddle point. Therefore, we need to add to the above answer  the contribution of the Andreev-Altschuler saddle point as well. Additionally, for the complete answer we also need to add the contribution of the $R^{\ddag}$, \eqref{eq.path.resolvents}. The computation of these terms closely follows equivalent computations presented in \cite{Altland:2021rqn}, and therefore we do not discuss these details here. Adding in all these contributions, one gets the full answer for the resolvent in the ergodic limit,
\begin{equation}
R(s)=	\frac{2\pi^2\rho^2(E)}{D^6} \left(\pi \delta(s) + 1 - \frac{\sin^2(s)}{s^2}\right) \tr \mathbb O \, \tr \mathbb O^\dagger ~.
\end{equation}
In this expression, note that the leading disconnected contribution corresponding to `1' is enhanced by a factor of $D^3$ with respect to the asymptotic contribution of the connected piece given by the $\delta(s)$. This happens because $s = \pi \rho(E) \omega \sim D^3 \omega$.

Let us consider only the connected contribution,
\begin{equation}\label{eq.connected-only}
	\frac{2\pi^2\rho^2(E)}{D^6} \left(\pi \delta(s) - \frac{\sin^2(s)}{s^2}\right) \tr \mathbb O \, \tr \mathbb O^\dagger ~.
\end{equation}
This term corresponds to the ramp-plateau behaviour depicted in \autoref{fig.g2t} after performing the Fourier transform to the time domain,
\begin{equation}
	\int({\rm connected\, term}) e^{i\omega t} = \left( 2 + t \, \frac{\Delta(E)}{\pi} + \left(2- t\,\frac{\Delta(E)}{\pi} \right)\, \text{sgn}\left(t\, \frac{\Delta(E)}{\pi} -2\right) \right)~.
\end{equation}
This term is zero at $t=0$. The disconnected term gives,
\begin{equation}
	\int({\rm disconnected\, term}) e^{i\omega t} = \frac{2\pi^2\rho^2(E)}{D^6} \tr \mathbb O \, \tr \mathbb O^\dagger \, \delta(t)~.
\end{equation}
The above integration gives rise to a Dirac-delta function only when the integration over $\omega$ is unbounded. In physical examples, the range of $\omega$ integration is fixed by the width of the spectrum ($\sim \sqrt{\bar a} \gamma$) and therefore the width of the {\it regulated}-delta function is controlled by $1/{(\sqrt{\bar a} \gamma)}$. Importantly, the value of the genus-2 partition spectral form factor is non-zero at $t=0$ (it is simply given by the square of the genus-2 partition function). 

From a gravitational point of view, there are two  types of geometry relevant for the genus-2 spectral form factor. Handlebody geometries \cite{Krasnov:2000zq}, which correspond to a bulk filling of a genus-2 surface. For the square of the genus-2 partition function, one can thus have two disconnected handlebodies. The second type of geometry is the genus-2 wormhole, see \cite{Maldacena:2004rf,Belin:2020hea}, which connects two genus-2 boundaries. One interesting aspect of the genus-2 wormhole is that it is a true saddle-point of the gravitational equations of motion, even at $t=0$. This should be contrasted with the double-trumpet or the double-cone \cite{Saad:2018bqo,Saad:2019lba,Cotler:2020ugk,Cotler:2020lxj} relevant for the usual spectral form factor. As we increase time, the genus-2 wormhole eventually dominates over the two disconnected handlebodies. It is not known when this happens exactly, but the fact that the genus-2 wormhole is a true saddle may play a role. In the current treatment, we have considered the homogeneous Hilbert space ansatz, which studies the theory only in the ergodic regime. At earlier times, it becomes crucial that the Hamiltonian is fixed. This can be captured by studying the contributions of the massive modes like in \cite{Altland:2021rqn}. It would be interesting to understand whether the fact that the genus-2 wormhole is a true saddle even at early times is important. We will address this in future work.


\section{Discussion}
\label{sec.discussion}
In this paper, we have analyzed the statistics of heavy-heavy-heavy OPE coefficients from the point of view of ergodicity and random matrix theory. To transform the problem into that of ordinary quantum mechanics, we defined a linearized operator on the tripled Hilbert space of the CFT whose matrix elements are given by the product of two OPE coefficients. We then proceeded to analyze the properties of this operator from a random matrix theory point of view. We argue that the behaviour of this operator in the ergodic limit can be explained by the statistics of the OPE coefficients, a conjecture originally proposed in \cite{Belin:2020hea}. Using the effective theory of quantum chaos, we also studied the genus-2 spectral form factor and found that it should have a ramp and plateau. 

\subsection{Open questions}

We conclude the paper with some open questions.

\subsubsection*{From CFT to QM} 

Part of the goal of this paper was to translate a CFT question which does not manifestly have a nice counterpart in quantum mechanics, into a framework that can be tackled strictly within quantum mechanics using random matrix theory. To do so, we defined a linear operator in a tripled Hilbert space. The next step was to assume that this operator is a random operator on the tripled Hilbert space. From the effective theory of quantum chaos, this amounts to assuming that the insertion of the operator source does not cause a large backreaction on the saddle-point. At this stage, this is an assumption, and it would be very interesting to understand whether this assumption is true. Also note that the framework we have presented would work for other tensors made out of more OPE coefficients, and it would be interesting to see whether the validity of the saddle-point assumption depends on the particular choice of tensor we made (i.e. on the particular combination of OPE coefficients).

This ties to a more profound understanding of what a heavy CFT operator is. An alternative route to that studied in this paper would be to study the expectation value of complicated (i.e. extensive) operators in high energy states \a`{a} la ETH. This could presumably be done numerically, although might quickly become difficult. The major problem that remains, even with the numerical methods, is to understand what the right basis of operators is. In CFTs, the state-operator correspondence tells us that local operators and energy eigenstates are one and the same. We don't see such an obvious choice in quantum mechanics. In a chaotic spin-chain, what is the ``right" extensive operator to pick? Should one consider randomized extensive operators?

Perhaps the most interesting theory to study in this regard is the SYK model. The SYK model can offer a bridge between a notion of complicated operator in QM and the CFT language, since the theory is conformal in the IR. A complicated operator would be one that is built from an $\mathcal{O}(1)$ fraction of the $N$ fermions, and the question becomes understanding what that maps to in the IR, in terms of the spectrum of local operators. This could be tackled using a combination of numerical and analytics techniques following up on \cite{Sonner:2017hxc,Nayak:2019khe,Nayak:2019evx}. 

Another point to mention is that CFTs always have more symmetries than generic quantum systems. One should thus organize the operators according to representations of the conformal group, and statements like the ETH should be always understood as applied to primary operators. This is particularly relevant in 2d CFTs, because of Virasoro symmetry. We do not expect the results to exhibit qualitative differences once this has been taken into account, but it be would be interesting to understand how this affects the EFT of quantum chaos.

\subsubsection*{Minimal models and the genus-2 spectral form factor}

Another interesting avenue to explore would be to study the genus-2 spectral form factor in rational CFTs that can be solved. The most natural example is the minimal models. For the usual spectral form factor, this was studied in \cite{Benjamin:2018kre}. Note that the minimal models are completely solved, so there is in no conceptual obstruction in computing the genus-2 spectral form factor. There are however some practical problems, coming from the fact that the genus-2 conformal blocks are not known in closed form, contrary to Virasoro characters. For plotting the genus-2 spectral form factor, this should not be too much of a problem since one can get the blocks numerically.

It would be interesting to check whether the minimal models at large level do display a ramp and plateau for the genus-2 spectral form factor. Even more interestingly, one could check if the dip time of the usual spectral form factor and that of its genus-2 counterpart are the same or not. In this paper, since we used only ergodicity arguments, we don't have an understanding of the Thouless time for the genus-2 spectral form factor. The Thouless time is both theory \textit{and} observable dependent, and therefore it is natural to think that the Thouless time could be quite different for the genus-2 spectral form factor.

\subsubsection*{The Symmetry Class of a CFT}
In section \ref{sec.ope}, we computed the mean and variance of the matrix elements of $\mathbb{O}$. We saw that the variance of the matrix elements involved a $\Tr \mathbb{O}^2$ term. The trace here is given by a particular cyclic contraction of the indices, and matches the decomposition of a genus-3 partition function in the sunset channel \cite{Belin:2021ryy}. However, we have obtained a single type of trace for $\mathbb{O}^2$ and there are in fact 5 possible trace structures involving four OPE coefficients (corresponding to the 5 OPE decompositions of a genus-3 surface), two examples of which are
\bea
\Tr_{\textrm{skyline}}c^4&=& c_{abc} c_{ade}^* c_{fbe} c_{fdc}^* \\
\Tr_{\textrm{comb}}c^4&=&c_{aad}c_{bbe}^*c_{ccf} c_{def}^* \,.
\eea
To correctly account for the statistics of the heavy-heavy-heavy OPE coefficients, we need all five of these structures \cite{Belin:2021ryy}. It is easy to see that the reason that we did not find all five structures lies in the way we applied unitary Haar averages to the operator $\mathbb{O}$.

One could imagine a more refined analysis in which we are more careful about the discrete symmetries of a CFT and the operator
under consideration. Generic CFT's have real and complex fields, and under complex conjugation of operators the OPE coefficients are also complex conjugated. This particular symmetry is not implemented in the unitary averaging we considered so
far, and taking it into account will presumably lead to a different type of averaging, a different type of RMT and and a different sigma model corresponding to a different symmetry class. It is our expectation that such a more precise treatment will lead to some quantitative changes in our result, in particular that a richer structure of random matrices and index structures will appear, but that the qualitative conclusions remain unaltered. Moreover, we expect that the leading order scaling with $D$ remains the same, and that we still
have a version of a ramp and a plateau in the relevant generalized form factors. We leave the interesting question which symmetry classes are realized in CFTs, with and without
operator statistics, and how these impact randomness and chaos, to a future study.

\subsubsection*{The OPE coefficients $c_{LLH}$}

There are essentially three types of OPE coefficients involving heavy operators, depending on whether one, two or three of the operators are heavy. As explained in the introduction, the operators $c_{LHH}$ are easiest to understand since the ETH ansatz makes a prediction for them. In this paper, we have developed a tool to study the OPE coefficients $c_{HHH}$. The final OPE coefficients to understand are thus $c_{LLH}$. Asymptotic formulas for such coefficients are also known \cite{Pappadopulo:2012jk,Das:2017cnv,Mukhametzhanov:2018zja,Collier:2019weq,6ptasymptotics} and it is natural to ask what random matrix theory has to say about them.

While our framework could in principle be used to study such operators as well, it seems important to underline a physical difference between these objects and the ones studied in this paper: there is no spectral form factor that one can define only with $c_{LLH}$ coefficients. In fact, the most Lorentzian probe of these OPE coefficients that one can cook up is simply the out-of-time-ordered four-point function on the plane. This is also a probe of quantum chaos, but of a quite different nature.\footnote{In fact, it really probes scrambling more than chaos. Even if the two are often related, they are not exactly the same thing. For example, the rate of scrambling is theory dependent while the eigenvalue statistics is much more universal. This suggests that how random the $c_{LLH}$ truly are could also be theory dependent. A similar observation was made in \cite{Caron-Huot:2020ouj}.} Since this object maps to a thermal correlator, but on Rindler space (a non-compact space), there is no discrete eigenspectrum to probe and hence no ramp and plateau. It would be interesting to understand better how random matrix theory constrains these OPE coefficients.


\section*{Acknowledgements}
We are happy to thank Alexander Altland, Daniel Jafferis, Diego Liska  for fruitful discussions. The presentation of certain results contained in this paper was improved following a comment posted by Brehm, Das and Datta. JdB is supported by the European Research Council
under the European Unions Seventh Framework Programme (FP7/2007-2013), ERC Grant
agreement ADG 834878.
JS thanks ENS Paris for hospitality during the final stages of this work.
This work has been partially supported by the SNF through Project Grants 200020 182513, as well as the NCCR 51NF40-141869 The Mathematics of Physics (SwissMAP).

\appendix
\section{Density of states in the tripled space} \label{appA}
In this appendix, we explore the relationship between the density of states of the tripled Hilbert space and that of the original Hilbert space.
When $\tau_i$s are all equal, the density of states in the tripled space is given by,
\begin{equation}\begin{aligned}
	\varrho( E) &= \lim_{\epsilon\to0} {\rm Tr} \left[ \frac \epsilon {(H_{\rm full}-z)^2+\epsilon^2 } \right] = \frac1{2\pi i} {\rm Tr}\left[ G^-(z) - G^+(z)\right],\\[5pt]
	&= \int dE_i \rho(E_1)\rho(E_2)\rho(E_3) \,\delta\! \left( E-\sum_i E_i\right) \\
	&= \sum_{n_1,n_2,n_3}\delta\!\left( E-\sum_i E_{n_i}\right)~.
\end{aligned}\end{equation}
This means that there is always 3-fold degeneracy. Evidently, the density of states in the tripled Hilbert space is a convolution of the density of states of the original Hilbert space.


\section{Haar integrals}
\subsection{Variance of the OPE coefficients corresponding to different ensembles}
\label{app.variance}
This section details the variance computation of the product of OPE coefficients leading to the result \eqref{eq.variance}. The variance we are interested in is given by the expression
\begin{equation}
	\overline{\langle i_i i_2 i_3|\mathbb O |i'_1 i'_2 i'_3\rangle^2}- \overline{\langle i_i i_2 i_3|\mathbb O |i'_1 i'_2 i'_3\rangle}^2~ \,.
\end{equation}
The first term of the above expression is given by,
\begin{align}
	\overline{\langle i_i i_2 i_3|\mathbb O |i'_1 i'_2 i'_3\rangle^2} &= \!\!\int\!\! \prod_{m=1}^3\!\!\left[dU^{(m)}\right] {} U_{i_1j_1}^{(1)} U_{i_2j_2}^{(2)} U_{i_3j_3}^{(3)} U_{j'_1i'_1}^{(1)\dagger} U_{j'_2i'_2}^{(2)\dagger} U_{j'_3i'_3}^{(3)\dagger} {} U_{i'_1k'_1}^{(1)} U_{i'_2k'_2}^{(2)} U_{i'_3k'_3}^{(3)} U_{k_1i_1}^{(1)\dagger} U_{k_2i_2}^{(2)\dagger} U_{k_3i_3}^{(3)\dagger}\nonumber \\
									  &\hspace{3cm} \times \Omega^*_{j_1j_2j_3} \Omega_{j'_1j'_2j'_3} \Omega^*_{k'_1k'_2k'_3}\Omega_{k_1k_2k_3} \nonumber\\
									  &= \prod_{m=1}^3 \bigg( \delta_{i_mi'_m}\Big(\delta_{j_mj'_m}\delta_{k_mk'_m} {\rm Wg}(D_m,2,\mathds 1) - \delta_{j_mk_m}\delta_{j'_mk'_m} {\rm Wg}(D_m,2,(2~1))\Big)\nonumber \\
									  &\hspace{2cm} + \Big(\delta_{j_mk_m}\delta_{j'_mk'_m} {\rm Wg}(D_m,2,\mathds 1) - \delta_{j_mj'_m}\delta_{k_mk'_m} {\rm Wg}(D_m,2,(2~1))\Big)\bigg) \nonumber \\
									  &\hspace{3cm} \times \Omega^*_{j_1j_2j_3} \Omega_{j'_1j'_2j'_3} \Omega^*_{k'_1k'_2k'_3}\Omega_{k_1k_2k_3}
\end{align}
There are a total of 64 terms in the above expression and let us look at only the leading order terms to begin with. Because of the following scaling property of the Weingarten functions,
\begin{equation}\label{eq.asympt.Wg}
	{\rm Wg}(D,n,\sigma) ~\substack{D\gg 1\\\sim} ~ D^{-n-|\sigma|}
\end{equation}
we need to only consider the terms proportional to ${\rm Wg}(D,2,\mathds 1)$. Here, $|\sigma|$ is the minimum number of pair exchanges required to build the permutation $\sigma$.
\begin{align}
	\overline{\langle i_i i_2 i_3|\mathbb O |i'_1 i'_2 i'_3\rangle^2} & \nonumber \\
	&\hspace{-1cm}\approx \prod_{i=1}^3{\rm Wg}(D_i,2,\mathds 1) \prod_{m=1}^3 \!\!\bigg( \delta_{i_mi'_m} \delta_{j_mj'_m}\delta_{k_mk'_m} + \delta_{j_mk_m}\delta_{j'_mk'_m} \bigg) \Omega^*_{j_1j_2j_3} \Omega_{j'_1j'_2j'_3} \Omega^*_{k'_1k'_2k'_3}\Omega_{k_1k_2k_3}\nonumber \\
									  &= \prod_{i=1}^3{\rm Wg}(D_i,2,\mathds 1)\bigg[ ~ \Omega_{j_1j_2j_3}\Omega^*_{j_1j_2j_3} ~ \Omega_{k_1k_2k_3}\Omega^*_{k_1k_2k_3} \Big(\delta_{EE'} + 1\Big) \nonumber \\
									  &\hspace{-1.9cm}+ ~ \Omega_{j_1j_2j_3}\Omega^*_{j_1j_2k_3} ~ \Omega_{k_1k_2k_3}\Omega^*_{k_1k_2j_3} \Big(\delta_{i_1i'_1}\delta_{i_2i'_2} + \delta_{i_1i'_1}\delta_{i_3i'_3} + \delta_{i_3i'_3}\delta_{i_2i'_2} ~+ \delta_{i_1i'_1} + \delta_{i_2i'_2} + \delta_{i_3i'_3}\Big) \bigg]
\end{align}
Subtracting off the disconnected contribution, \eqref{eq.signal-diff}, we get,
\begin{align}\label{eq.variance.diff-app}
	\overline{\langle i_i i_2 i_3|\mathbb O |i'_1 i'_2 i'_3\rangle^2}- \overline{\langle i_i i_2 i_3|\mathbb O |i'_1 i'_2 i'_3\rangle}^2&\nonumber\\
																	    &\hspace{-3cm} \approx \left(\prod_{i=1}^3{\rm Wg}(D_i,2,\mathds 1) - \prod_{i=1}^3{\rm Wg}(D_i,1,\mathds 1)^2\right) ~ \Omega_{j_1j_2j_3}\Omega^*_{j_1j_2j_3} ~ \Omega_{k_1k_2k_3}\Omega^*_{k_1k_2k_3} \delta_{EE'}\nonumber \\
																	    &\hspace{-4.5cm}+ \prod_{i=1}^3{\rm Wg}(D_i,2,\mathds 1)\bigg[ ~ \Omega_{j_1j_2j_3}\Omega^*_{j_1j_2j_3} ~ \Omega_{k_1k_2k_3}\Omega^*_{k_1k_2k_3}  \nonumber \\
									  &\hspace{-4.8cm}+ ~ \Omega_{j_1j_2j_3}\Omega^*_{j_1j_2k_3} ~ \Omega_{k_1k_2k_3}\Omega^*_{k_1k_2j_3} \Big(\delta_{i_1i'_1}\delta_{i_2i'_2} + \delta_{i_1i'_1}\delta_{i_3i'_3} + \delta_{i_3i'_3}\delta_{i_2i'_2} ~+ \delta_{i_1i'_1} + \delta_{i_2i'_2} + \delta_{i_3i'_3}\Big) \bigg]
\end{align}
Using the following definitions for the Weingarten functions,
\begin{equation}
	{\rm Wg}(D,2,\mathds 1) = \frac1{D^2-1}, \quad {\rm Wg}(D,1,\mathds 1) = \frac1D~,
\end{equation}
we get (treating the $D_i\sim D$),
\begin{equation}
	{\rm Wg}(D,2,\mathds1)^3 - {\rm Wg}(D,1,\mathds1)^6 = \frac{3D^4 -3D^2 +1}{D^6 (D^2-1)^3} ~ \substack{D\gg1\\=} ~ \frac3{D^8} + \cdots~.
\end{equation}
We see that the term in the first line of the RHS in \eqref{eq.variance.diff-app} is subleading. Also note that the terms in the last line on the RHS contribute only to a parametrically smaller number of elements of the operator, $\mathbb O$, these are ``partially off-diagonal" terms. The coefficient is given by
\be
\sum_{j_1,j_2,j_3,k_1,k_2,k_3}\Omega_{j_1j_2j_3}\Omega^*_{j_1j_2k_3} ~ \Omega_{k_1k_2k_3}\Omega^*_{k_1k_2j_3} \,.
\ee
This is again a trace, which means it is basis independent. We call such a term $\trt \mathbb{O}^2$ and we have
\be
\trt \mathbb{O}^2= C_{ijk}C^*_{ijl}C_{mnl}C_{mnk}^* \,.
\ee
This combination of OPE coefficients appears in the sunset decomposition of the genus-2 partition function \cite{Belin:2021ryy}. Therefore, we call this term $\trt_{\rm sunset} \mathbb{O}^2$.

\subsection{Variance of the OPE coefficients corresponding to identical ensembles} \label{haarsame}
This section details the variance computation of the product of OPE coefficients leading to the result, \eqref{eq.variance.same}. The variance we are interested in is given by the expression,
\begin{equation}
	\overline{\langle i_i i_2 i_3|\mathbb O |i'_1 i'_2 i'_3\rangle^2}- \overline{\langle i_i i_2 i_3|\mathbb O |i'_1 i'_2 i'_3\rangle}^2~,
\end{equation}
but with all the states lying in the same microcanonical ensemble this time. The first term of the above expression is given by,
\begin{align}
	\overline{\langle i_i i_2 i_3|\mathbb O |i'_1 i'_2 i'_3\rangle^2} &=\!\! \int\!\! \prod_{m=1}^3\!\!\left[dU^{(m)}\right] {} U_{i_1j_1} U_{i_2j_2} U_{i_3j_3} U_{i'_1j_4} U_{i'_2j_5} U_{i'_3j_6} U_{j'_1i'_1}^{\dagger} U_{j'_2i'_2}^{\dagger} U_{j'_3i'_3}^{\dagger} {} U_{j'_4i_1}^{\dagger} U_{j'_5i_2}^{\dagger} U_{j'_6i_3}^{\dagger}\nonumber \\
									  &\hspace{3cm} \times \Omega^*_{j_1j_2j_3} \Omega_{j'_1j'_2j'_3} \Omega^*_{j_4j_5j_6}\Omega_{j'_4j'_5j'_6} \nonumber\\
									  &= \sum_{\sigma,\tau\in S_6} {\rm Wg}(D_1,6,\sigma \tau^{-1}) \prod_{k=1}^6 \delta_{i_ki'_{\sigma(k)}} \delta_{j_kj'_{\tau(k)}}
\end{align}
There are a total of $6!^2$ terms in this expression, but at the leading order only terms where $\tau = \sigma$ contribute (see \eqref{eq.asympt.Wg}). Their contribution is proportional to ${\rm Wg}(D,6,\mathds 1) \sim 1/D^{6}$. This is still a large number of terms ($720$). But, a smaller subgoup, $S_3 \times S_3$, only generates the terms that permute the three individual states defining the full state, $|i_1 i_2 i_3\rangle$. As we mentioned in the main text, when all states belong to the same microcanonical window, then this difference is superficial and doesn't give rise to new terms. Discounting this redundancy, we are left with only 20 different terms. Only one of these terms corresponds to the disconnected piece.
Subtracting off the disconnected contribution, \eqref{eq.signal-same}, we get,
\begin{align}\label{eq.variance-app}
	\overline{\langle i_i i_2 i_3|\mathbb O |i'_1 i'_2 i'_3\rangle^2}- \overline{\langle i_i i_2 i_3|\mathbb O |i'_1 i'_2 i'_3\rangle}^2&\nonumber\\
																	    &\hspace{-4cm}\approx \left({\rm Wg}(D_1,6,\mathds 1) - \frac1{D_1^2(D_1+1)^2(D_1+2)^2}\right) ~ \Omega_{j_1j_2j_3}\Omega^*_{j_1j_2j_3} ~ \Omega_{k_1k_2k_3}\Omega^*_{k_1k_2k_3} [\delta_{EE'}]_{S_3}\nonumber \\
																	    &\hspace{-6cm}+ {\rm Wg}(D_1,6,\mathds 1)\bigg[ ~ \Omega_{j_1j_2j_3}\Omega^*_{j_1j_2j_3} ~ \Omega_{k_1k_2k_3}\Omega^*_{k_1k_2k_3} \left( 1+\delta_{i_1i_2} + \delta_{i_2i_3} + \delta_{i_3i_1} + \delta_{i_1i_2} \delta_{i_2i_3} \delta_{i_3i_1} \right) \times (\cdots)'  \nonumber \\
									  &\hspace{-4.8cm}+ ~ \Omega_{j_1j_2j_3}\Omega^*_{j_1j_2k_3} ~ \Omega_{k_1k_2k_3}\Omega^*_{k_1k_2j_3} \Big(\delta_{i_1i'_1}\delta_{i_2i'_2} + \cdots 647 \text{ terms } \cdots \Big) \bigg]~.
\end{align}
On the right hand side, in the second line $\left(\cdots\right)'$ represents the terms in the first bracket but with the primed indices. In the third line the 648 terms correspond to partial contractions between the primed and the unprimed indices corresponding to the external bra-ket states.

Once again, using the asymptotic properties of the Weingarten functions we get,
\begin{equation}
	{\rm Wg}(D_1,6,\mathds1) - \frac1{D_1^2(D_1+1)^2(D_1+2)^2} \sim \mathcal O(D_1^{-7}).
\end{equation}
We see that the term in the first line of the RHS in \eqref{eq.variance-app} is subleading.
\bibliographystyle{utphys}
\bibliography{ref}
\end{document}